\documentclass[12pt]{article}

\begin{document}

\def\noa#1{\noalign{\vskip#1pt}}

\begin{center}
{\Large\textbf{Triality and Dual Equivalence Between Dirac\\[0.2cm]
Field and Topologically Massive Gauge Field}}\\[0.6cm]

Liu Yu-Fen  \\[0.2cm]
\textit{\small Institute of Theoretical Physics, Academia Sinica, Beijing
100080, China}\\
email: liuyf@itp.ac.cn \\
( 2006. February.)
\end{center}

\begin{abstract}
It is proved that there exist a vector representation of Dirac's
spinor field and in one sense it is equivalent to biquaternion
(i.e.\ complexified quaternion) representation. This can be
considered as a generalization of Cartan's idea of triality to
Dirac's spinors. In the vector representation the first order
Dirac Lagrangian is dual equivalent to the two order Lagrangian of
topologically massive gauge field. The potential field which
corresponds to the Dirac field is obtained by using master (or
parent) action approach. The novel gauge field is self-dual and
contains both anti-symmetric Lie and symmetric Jordan structure.
\end{abstract}

\vspace*{0.3cm}

\section{Introduction}

\vspace{1pt} Dualization has appeared in several different
contexts in theoretical physics. Here  \textbf{duality } will mean
that there exist two equivalent descriptions of a model using
different fields. A classical example is the fermion-boson duality
between the massive Thirring model for spinor $S(\psi)$ and
sine-Gordon model for boson $S(\phi)$. Up to date only in certain
2D-dualities do we have an explicit relation between $S(\psi)$ and
$S(\phi)$ where $\phi\sim\bar{\psi}\psi$ is a bound state from the
point of view of Thirring model$^{\lbrack 1\rbrack }$. Techniques
such as ``bosonization" showed equivalence between theories of
drastically different appearance. The boson in one approach might
appear as a bound state of fermions in another, but in terms of
the respective Lagrangian approaches, they were equally
fundamental. Many phenomena that are difficult to understand in
the fermion language has simple semiclassical explanation in the
Bose language. Another classical example is the duality between
the first order 3D massive self-dual vector field and two-order 3D
topologically massive gauge field$^{\lbrack 2, 3\rbrack}$. It is
known (up to date) that only in three dimensions, a topological
Chern--Simons term (which must includes 3D Levi-Civita symbol) is
bilinear and can produce a mass. The study of such a gauge theory
has particularly interest because the Chern--Simons
characteristics supply a gauge-invariant mass-generating
mechanism, and novel topological mass term is available.

In this paper we are interested in the vector representation of
Dirac's spinors in (3+1) dimensional Minkowski space
(spinor-vector duality) and the mass-generating mechanism for
Dirac's particles.

The first example of vector representation for spinors was derived
by E.\ Cartan$^{\lbrack 4\rbrack }$ for SO(8) case. Spinors can be
decomposed (under the group of rotations) into \textit{two}
irreducible parts, which are the two types of semi-spinors. In
general, the dimension of (real Euclidian) vector space $D_{V}=n$
(even) is not equal to the dimension of the corresponding
semi-spinor space of the first type $D_{S1}=2^{n/2-1}$ and
dimension of semi-spinor space
 of the second type $D_{S2}=2^{n/2-1}$.
But in the special case, \textbf{only} when $n=8$, the above
\textit{three} spaces have the \textit{same} dimension (i.e
$D_{V}=D_{S1}=D_{S2}=8$), all real and are remarkably on an equal
footing. In fact, the vector representation of above two
semi-spinors is obtained only in this special 8D case. The
interesting ``triality'', or symmetry between the above three
eight-dimensional spaces appears in this case.

The word ``\textbf{triality}'' was chosen by analogy with the word
``duality'', and is applied by Cartan  to the algebraic and
geometric aspects of the $\Sigma_{3}$ symmetry which Spin(8) has.
More general notion of triality is due to Adams$^{[5, 6,7]}$. One
needs to know what three objects the symmetry group $\Sigma _{3}$
permute; and the answer is that it permutes
\textbf{representations}, i.e.\ triality permutes vector and two
semi-spinors representations. Moreover, the ``8-squares formula"
in the real domain (i.e.\ the product of two sums of eight squares
is itself the sum of eight squares.) which deduced from the
triality formulation by Cartan$^{\lbrack 4\rbrack }$, can be
considered as a \textit{normed} condition for division
\textit{octonion}$^{\lbrack 8\rbrack }$. It means that for the
above SO(8) case, the vector representation of the semi-spinors is
equivalent to octonion representation! (Historically, in 1845 an
``8-squares formula" was discovered by A. Cayley ; J.T. Graves
claimed to have discovered this earlier, and in fact C.F. Degan
had already noted a more general formula in 1818.)

By the Hurwitz$^{\lbrack 9\rbrack}$ and Frobenius$^{\lbrack
10\rbrack} $ theorems, the only finite dimensional division
algebras over reals are: \textbf{R} (real), \textbf{C} (complex),
\textbf{H} (quaternion), \textbf{O} (octonion) algebras. Any
division algebra gives a triality, and it follows that
\textit{normed} trialities only occur in dimensions 1, 2, 4 or 8.
Here we use a generalized concept of ``triality" advocated by
Adams$^{[5,6,7]}$. This conclusion is quite deep. By comparison,
Hurwitz's classification of \textit{normed} division algebra (in
real domain) is easier to prove. It is important to stress that
the concept of ``triality" comes from the theory of vector
representation of two semi-spinors. In our theory ``triality" can
be considered as an algebraic properties of biquaternion algebra,
which was needed in the vector reformulation of Dirac's spinors.

We propose here an alternative way to generalize Cartan's idea to
the Dirac's spinors which associated with Minkowski space. We will
prove that there exist a vector representation of Dirac's spinor
field, and in one sense it is equivalent to biquaternion (i.e.\
complexified quaternion) representation. The symbol $\check{c}
^{\mu \nu \lambda }$ which characterizes the bilinear law of
composition for biquaternion (and plays crucial role in our
theory), contains both (anti-symmetric) Lee and (symmetric) Jordan
structure and this definition preserve Lorenz invariance$^{\lbrack
11\rbrack }$. Furthermore we will prove that in vector
representation the first order Dirac Lagrangian is dual-equivalent
to the two order Lagrangian of the topologically massive self-dual
gauge field. Here we will use parent (or master) action approach
which was proposed by Deser and Jackiw$^{\lbrack 2\rbrack }$. This
approach was in fact originated from the Legendre transformation,
but an explicit description of one as a function of the other
would be non-local and non-linear. The potential (gauge) field
which corresponds to Dirac's vector field is obtained by using
this approach.

\section{The algebra of biquaternion}

\vspace{1pt}The quaternion was discovered by W.R.\ Hamilton in
1843. He regarded it as his most important contribution and
labored the rest of his life, for 22 years, to formulate
everything in physics and astronomy in terms of quaternion. But he
was not successful. In a letter to a friend in 1844 he wondered
whether the vector part could represent the three space dimensions
and the scalar part represent time! Since the field of quaternion
is in a 4-dimensional Euclidean space, \textbf{complex} components
for quaternions are required for $1+3$ space-time of special
relativity $^{\lbrack 12\rbrack}$. Quaternions of this nature were
called \textbf{biquaternion} by Hamilton, and do not form a
division algebra.

\vspace{1pt}Let $G^{\mu }$, $H^{\nu }$, $S^{\lambda }$ (here $\mu
=1,2,3,\ldots,n$) are vectors. The dot product (scalar product)
and bilinear law of composition for vectors (tensor product) are
defined here as
\begin{equation} %(1)
\begin{array}{rll}
G\cdot H & \stackrel{\mbox{\scriptsize def.}}{\longrightarrow} & G^{\mu }g_{\mu \nu }H^{\nu } \\
(G\otimes H) & \stackrel{\mbox{\scriptsize def.}}{\longrightarrow}
& G_{\mu }c^{\mu \lambda \nu }H_{\nu }=S^{\lambda }
\end{array}
\end{equation}
The metric $g_{\mu \nu }$ and the symbol $c^{\mu \lambda \nu }$ which
satisfy the following ``normed condition''
\begin{equation} % (2)
(G\cdot G)(H\cdot H)=(G\otimes H)\cdot (G\otimes H)
\end{equation}
completely define the structure of the algebra, which is needed
here. (Here we shall use the convention: the word ``algebra" will
mean ``algebra with unit element".)

In 1898 Hurwitz$^{\lbrack 9\rbrack }$ proved that for real
vectors, if $g_{\mu \nu }=\delta _{\mu \nu } $\ (Euclidean) then
the above algebraic identities are possible \textit{only} for
$n=1,2,4,8$. This result, of course, is intimately related to the
well-known result, due to $\cdots$ that there exist only four
normed algebras over reals: \textbf{R} (real), \textbf{C}
(complex), \textbf{H} (quaternion), \textbf{O} (octonion). Each
algebra has unit element $k^{\mu }$ (real) such that for any
(real)  $G^{\mu }$
\begin{equation} %(3)
\begin{array}{lll}
k_{\mu }c^{\mu \lambda \nu }G_{\nu }=G_{\mu }c^{\mu \lambda \nu
}k_{\nu }=G^{\lambda } & , & k^{\mu }\delta _{\mu \nu }k^{\nu }=1
\end{array}
\end{equation}
The unit element is necessarily unique and characterize the
representation of the algebra. From abstract point of view the
metric $g_{\mu \nu }$ and the symbol $c^{\mu \lambda \nu }$
completely define the structure of the algebra and the unit vector
$k^{\mu }$ is ``hidden inside" $c^{\mu \lambda \nu }$.

We must distinguish between the composition which denoted here by
symbol  $c^{\mu \lambda \nu }$ and the antisymmetric cross product
which is used in Maxwell theory of electromagnetism. As is well
known, the natural generalization of the  concept of the
antisymmetric bilinear cross product (and curl) in higher
dimensional space is possible only in $n=1,3,7$ dimensional
spaces$^{[13]}$, and they come from the \textbf{C}, \textbf{H},
\textbf{O}. For physicist  it means that the natural
generalization of Maxwell theory of electromagnetism in higher
dimensional space possible only in seven dimensional space. In
other words the construction of totally antisymmetric three index
symbol $\epsilon ^{\mu \nu \lambda }$ possible only in 3 and 7
dimensions. (Comments: The above result is a good reason why $S^2$
and $S^6$ are the only spheres that can have an almost complex
structure. In the case of $S^2$ the almost complex structure comes
from an honest complex structure on the Riemann sphere. The
6-sphere, $S^6$, when considered as the set of unit norm imaginary
octonion, inherits an almost complex structure from the octonion
multiplication. Recently Chern proposed (``last Chern theorem")
that none of the almost complex structure on $S^6$ can be
integrable and argued that from this it follows that $S^6$ does
not admit any complex structure$^{[14]}$.)

In fact, our physical space is 4-dim. Minkowski space with
signature $(+,-,-,-)$. Moreover, there is the \textit{complex}
continuum of quantum mechanics, which gives rise to the concept of
probability amplitude and superposition law, leading to the
picture of \textbf{complex} Hilbert space for description of
phenomena. Thus here we need the concept of (complexified)
biquaternion. (Notice that the \textit{dimension} and
\textit{signature} of space-time are now necessarily fixed; a
biquaternion description would not work if these did not have the
unique values that we observe for our physical universe.) For
these, let $G^{\mu }$, $H^{\nu }$, $S^{\lambda }$ are complex
vectors and
\begin{equation} %(4)
\begin{tabular}{l}
$g_{\mu \nu }=\eta _{\mu \nu }=\mbox{diag}(1,-1,-1,-1)$ \\\noa3
$c^{\mu \nu \lambda }\stackrel{\mbox{\scriptsize def.}}{=} (t^{\mu
\nu \lambda \rho }-i\epsilon ^{\mu \nu \lambda \rho })k_{\rho
}=t^{\mu \nu \lambda }-i\epsilon ^{\mu \nu \lambda }$
\end{tabular}
\end{equation}
here $\epsilon ^{\mu \nu \lambda \rho }$ is the Levi-Civita
symbol, with $\epsilon ^{0123}=1$.
\begin{equation} %(5)
\begin{array}{l}
k^{\mu }\eta _{\mu \nu }k^{\nu }=1 \\\noa4 \epsilon ^{\mu \nu
\lambda \rho} =\frac{i}{4} tr(\gamma ^{5}\gamma
^{\mu }\gamma ^{\nu }\gamma ^{\lambda }\gamma ^{\rho }) \\
\noa6 t^{\mu \nu \lambda \rho }\stackrel{\mbox{\scriptsize
def.}}{=}\frac{1}{4}tr(\gamma ^{\mu }\gamma ^{\nu }\gamma
^{\lambda }\gamma ^{\rho })=(\eta ^{\mu \nu }\eta ^{\lambda \rho
}+\eta ^{\mu \rho }\eta ^{\nu \lambda }-\eta ^{\mu \lambda }\eta
^{\nu \rho })  \\\noa6 \displaystyle c^{\mu \nu \lambda
\rho}\stackrel{\mbox{\scriptsize def.}}{=}(t^{\mu \nu \lambda \rho
}-i\epsilon ^{\mu \nu \lambda \rho })
\end{array}
\end{equation}
and $\gamma ^{\mu}$ are Dirac's matrices. We can prove that the above
$\eta _{\mu \nu }$ and $c^{\mu \nu \lambda}$ satisfy Eq.\ (2)
and
\begin{equation} %(6)
c^{\mu \nu \sigma }\eta _{\sigma \delta }
(c^{\delta \rho \lambda })^{\ast
}+c^{\mu \rho \sigma }\eta _{\sigma \delta }
(c^{\delta \nu \lambda })^{\ast
}=2\eta ^{\mu \lambda }\eta ^{\nu \rho }
\end{equation}
in which asterisk $\ast $ represents complex conjugate and
$(c^{\delta \rho \lambda })^{\ast }=c^{\lambda \rho \delta}$.
(Raising and lowering indices by Lorentzial metric $\eta ^{\mu
\nu}$ and $\eta _{\mu \nu}$, the summation convention is assumed
for repeated indices.) These relations define a structure of the
algebra of biquaternion $^{[11,12]}$.
The unit (time-like real) vector $k^{\mu }$ means a preferred
direction (i.e.\ the ``time"-direction) in Minkowski space-time, or
a split of the physical space-time into space and time.

Because the above algebra is associative, it can be considered in
terms of the matrices (which are more familiar to physicist). Let
the matrices $e^\mu $ be the hypercomplex basic units, the rows
and columns of matrices $e^\mu $ are labelled with four-component
indices $\nu ,\lambda$ etc., with $(e^\mu )_{\cdot \lambda }^{\nu
\cdot }$. The matrix representation of biquaternion can be
expressed as
\begin{equation} %(7)
\begin{array}{ccc}
G=G_\mu e^\mu & ; & (e^\mu )_{\cdot \lambda }^{\nu \cdot }
\stackrel{\mbox{\scriptsize def.}}{=}%
c^{\nu \sigma \mu }\eta _{\sigma \lambda }
\end{array}
\end{equation}
here $G_\mu $ are complex. We can prove that a bilinear law of
composition (1) is equivalent to matrix multiplication $(G_\mu
e^\mu )_{\cdot \sigma }^{\lambda \cdot}(H_\nu e^\nu )_{\cdot
\rho}^{\sigma \cdot} =(S_\mu e^\mu )_{\cdot \rho }^{\lambda \cdot
}$. In the special coordinate system, where $k^\mu =\delta ^{\mu
0}$, the basis element $e^0=I$ and for $\nu =1,2,3$ we have $e^\nu
=\sqrt{-1}\hat{e}^\nu $ here $\hat{e}^\nu $ obey the
multiplication law:
\begin{equation} %(8)
\begin{array}{c}
\hat{e}^1\hat{e}^1=\hat{e}^2\hat{e}^2=\hat{e}^3\hat{e}^3=
\hat{e}^1\hat{e}^2 \hat{e}^3=-I \\\noa3
\hat{e}^1\hat{e}^2=-\hat{e}^2\hat{e}^1=\hat{e}^3
\end{array}
\end{equation}
(the last relation being cycle). These relations characterize the
algebra of \textit{complexified} \textit{bi}quaternions$^{[12]}$.
Although the matrix representation of the biquaternion algebra is
more familiar to physicist, but I do not use it here. It is
convenient for me to work with a tensor $c^{\mu \nu \lambda}$. In
the tensor language, the second-rank tensor $(c^{\nu \sigma \mu
}G_\mu)^{\ast}$ (which corresponds to matrix
$(e^{\mu}G_{\mu})^\ast )$ is self-dual (see Eqs.\ (45)--(47)
below).

In Minkowski space, the dot product for complex vectors is not
necessarily real, let alone positive. For some physical
considerations, instead of time-like (real) unit vector  $k^{\mu
}$ we can introduce space-like (real) unit vector  $j^{\mu }$ such
that $j^{\mu }\eta _{\mu \nu }k^{\nu }=0$ and instead of $c^{\mu
\nu \lambda}$ (which defined in Eqs.\ $(4)\sim(6)$) we introduce
$\check{c}^{\mu \nu \lambda}$ such that
\begin{equation} %(9)
\begin{array}{c}
j^{\mu }\eta _{\mu \nu }j^{\nu }=-1 \\
\noa4 \check{c}^{\mu \nu \lambda }\stackrel{\mbox{\scriptsize
def.}} {=}(t^{\mu \nu \lambda \rho }-i\epsilon ^{\mu \nu \lambda
\rho })j_{\rho }=\check{t}^{\mu \nu \lambda }-
i\check{\epsilon} ^{\mu \nu \lambda } \\
\noa4 j\check{\otimes}G\stackrel{\mbox{\scriptsize def.}}
{\longrightarrow}j_{\mu }\check{c}^{\mu \lambda \nu }G_{\nu }
=-G^{\lambda }
\\\noa4 (G\cdot G)(H\cdot
H)=-(G\check{\otimes}H)\cdot (G\check{\otimes}H) \\
\noa4 \check{c}^{\mu \nu \sigma }\eta _{\sigma \delta
}(\check{c}^{\delta \rho \lambda })^{\ast } +\check{c}^{\mu \rho
\sigma }\eta _{\sigma \delta }(\check{ c}^{\delta \nu \lambda
})^{\ast }=-2\eta ^{\mu \lambda }\eta ^{\nu \rho }
\end{array}
\end{equation}
On the analogy of Dirac's ``gama five'' matrix $\gamma_{5}$, we
introduce unipotent $ c_{5}^{\mu \nu }$ such that
\begin{equation} %(10)
\begin{array}{c}
c_{5}^{\mu \lambda }\stackrel{\mbox{\scriptsize def.}}{=}j_{\nu
}c^{\nu \mu \lambda \rho}k_{\rho}=k_{\nu }\check{c}^{\nu \lambda
\mu } \\\noa4 c_{5}^{\mu \rho }\eta _{\rho \sigma }c_{5}^{\sigma
\nu }=\eta ^{\mu \nu } \\\noa4 \check{c}^{\mu \nu \lambda
}=-c^{\mu \nu \sigma }\eta _{\sigma \rho }c_{5}^{\rho \lambda
}=(c_{5}^{\mu \sigma })^{\ast }\eta _{\sigma \rho }c^{\rho \nu
\lambda }
\end{array}
\end{equation}
The last relation can be considered as the mapping from ${c}^{\mu
\nu \lambda } \rightarrow \check{c}^{\mu \nu \lambda }$. We see
that ($\eta _{\mu \nu }$, $\check{c}^{\mu \lambda \nu }$) can be
considered as another representation of the algebra of
biquaternion which equivalent to ($\eta _{\mu \nu }$, $c^{\mu
\lambda \nu }$). For some physical consideration, we will work
only with $\eta _{\mu \nu }$ and $\check{c}^{\mu \lambda \nu }$
here.

The Dirac operator is defined as
\begin{equation} %(11)
\begin{array}{lll}
\check{D}^{\mu \nu}\stackrel{\mbox{\scriptsize def.}}{=}
\check{c}^{\mu \lambda \nu}\partial_{\lambda} & , &
(\check{D}^{\mu \lambda})^{\ast } \eta _{\lambda \rho
}(\check{D}^{\rho \nu})=- \eta ^{\mu \nu }\partial\partial
\end{array}
\end{equation}

Any 4-dimensional biquaternion $G_{\mu }$ can be decomposed in the
sum of two parts:  vector part and scalar part. For this, without
lose of generality, we introduce unipotent $c_{3}^{\mu \nu }
\stackrel{\mbox{\scriptsize def.}}{=}\check{c}^{\mu \lambda \nu }
j_{\lambda}$ (which has property $c_{3}c_{3}=I$) and two mutually
annihilating idempotents
\begin{equation} %(12)
\begin{array}{l}
p_{+}^{\mu \nu }\stackrel{\mbox{\scriptsize def.}}{=} \frac{1}{2}
(\eta ^{\mu \nu }+c_{3}^{\mu \nu })=\eta^{\mu \nu
}+j^{\mu}j^{\nu}\\ \noa6 p_{-}^{\mu \nu
}\stackrel{\mbox{\scriptsize def.}}{=} \frac{1}{2}(\eta ^{\mu \nu
}- c_{3}^{\mu \nu })=-j^{\mu}j^{\nu} \\
G^{\mu }=p_{+}^{\mu \nu }G_{\nu}+p_{-}^{\mu \nu }
G_{\nu}=\mathcal{G}^{\mu}+(-j^{\mu}\phi)
\end{array}
\end{equation}
Here $\mathcal{G}^{\mu} \stackrel{\mbox{\scriptsize def.}}{=}
p_{+}^{\mu \nu }G_{\nu}$ is the vector part of biquaternion (which
has only three independent complex components) and $\phi
\stackrel{\mbox{\scriptsize def.}}{=}G^{\nu}j_{\nu}$ is the
complex scalar.

The \textit{associative} biquaternion product $G\check{\otimes} K
\rightarrow G_\mu (\check{t}^{\mu \nu \lambda }- i\check{\epsilon}
^{\mu \nu \lambda }) K_\lambda$ can be decomposed into
antisymmetric and symmetric parts. The antisymmetric part
$\frac{1}{2}(G\check{\otimes} K-K\check{\otimes}
G)\stackrel{\mbox{\scriptsize def.}} {\longrightarrow} -i
G_{\mu}\check{\epsilon}^{\mu \nu \lambda }K_{\lambda}$ corresponds
to the Lie algebra which generates generalized rotations. There is
Jacobi identity
\begin{equation} %(13)
G_{\mu}\check{\epsilon}^{\mu \nu \lambda
}[K_{\rho}\check{\epsilon}^{\rho \lambda \gamma}H_\gamma ]
+K_{\mu}\check{\epsilon}^{\mu \nu \lambda
}[H_{\rho}\check{\epsilon}^{\rho \lambda \gamma}G_\gamma ]
+H_{\mu}\check{\epsilon}^{\mu \nu \lambda
}[G_{\rho}\check{\epsilon}^{\rho \lambda \gamma}K_\gamma ]=0
\end{equation}
What about the symmetric part of the product? It is
\begin{equation} %(14)
(G{\circ} K) \stackrel{\mbox{\scriptsize def.}}{=} \frac{1}{2}
(G\check{\otimes} K+K\check{\otimes} G)
\stackrel{\mbox{\scriptsize def.}}
{\longrightarrow}G_{\mu}\check{t}^{\mu \nu \lambda }K_{\lambda}
\end{equation}
That is \textit{non-associative} special Jordan algebra product.
There are Jordan identities
\begin{equation} %(15)
\begin{array}{c}
G{\circ} K=K{\circ} G \\\noa3 ((G{\circ} G){\circ} K){\circ}
G=(G{\circ} G){\circ}(K{\circ} G)
\end{array}
\end{equation}
Roughly speaking, the symmetric Jordan algebra generates
generalized radial distortion. (Comments: Jordan algebra invented
as a way to study the self-adjoint operators on a Hilbert space,
which represent observables in Quantum mechanics. If you multiply
two self-adjoint operators the result needn't be self-adjoint, but
the above Jordan product is self-adjoint.)

Elements of non-associative Jordan algebra form a so-called Jordan
triple system under the Jordan triple product
\begin{equation} %(16)
\{ GKH \} \stackrel{\mbox{\scriptsize def.}}{=}
G{\circ}(K{\circ}H)+(G{\circ}K){\circ}H-K{\circ}(G{\circ}H)
\end{equation}
This triple product satisfies
\begin{equation} %(17)
\begin{array}{c}
\{ GKH \} =\{ HKG \} \\\noa3
\{ GK \{ HNB \} \} -\{ HN \{ GKB \} \} -\{ G \{ NHK \} B \} + \{
\{ HNG \} KB \} =0
\end{array}
\end{equation}
which are the defining identities of a Jordan triple system.

\section{Triality and vector representation of Dirac spinor}

\vspace{1pt} The dimension of spinor space, depends on both the
\textit{dimensions} of vector space and the \textit{signature} of
the metric. In 4-Lorentzian dimensions, the gamma matrices
$\gamma^\mu$ are (at least) $4\times4$, and thus the number of
complex spinor components must be four. According E.\ Cartan,
spinors can be decomposed, under the group of rotations, into
\textit{two} irreducible parts. That is any complex Dirac spinor
can be decomposed into the sum of two `semi-spinors'
$\Psi^{\alpha}=\Psi^{\alpha} _{1}+\Psi^{\alpha} _{2}$ (here
$\alpha $ is a spinor's index). Each semi-spinor has 4 independent
real components. Thus we have here \textbf{three} spaces each of
four dimensions, that of Lorentzian vectors, that of semi-spinors
of the first type, that of semi-spinors of the second type.
Corresponding to them we introduce unit-basis $(k^{\mu },j^{\nu
},u^{\beta },v^{\alpha})$, where $k^{\mu }$ and $j^{\nu }$ are the
time-like and space-like unit-vectors. The basic unit Dirac
semi-spinors $u^{\alpha }$ and $v^{\alpha }$ are defined such that
\begin{equation} %(18)
\begin{array}{lll}
k_{\nu}k^{\nu}=-j_{\mu}j^{\mu}=1  & ; & k_{\nu}j^{\nu}=0 \\\noa2
\bar{u}u=-\bar{v}v=1 & ; & \bar{u}v=0 \\\noa2
u=j_{\mu }i\gamma ^{\mu }v & ; & v=j_{\mu }i\gamma ^{\mu }u \\\noa2
v=-k_{\mu }\gamma ^{\mu }v & ; & u=k_{\mu }\gamma ^{\mu }u
\end{array}
\end{equation}
here $\bar{u}=u^{\dagger} \gamma^0$ represents Dirac
conjugate of $u$. Moreover
\begin{equation} %(19)
\begin{array}{ccc}
\bar{u}\gamma ^{\mu }\gamma ^{\nu }\gamma ^{\lambda }u & =\bar{v}%
\gamma ^{\mu }\gamma ^{\nu }\gamma ^{\lambda }v & =i\epsilon ^{\mu \nu
\lambda \rho }j_{\rho }+t^{\mu \nu \lambda \rho }k_{\rho } \\\noa2
-\bar{u}\gamma ^{\lambda }\gamma ^{\nu }\gamma ^{\mu }v & =\bar{v}%
\gamma ^{\mu }\gamma ^{\nu }\gamma ^{\lambda }u & =\epsilon ^{\mu \nu
\lambda \rho }k_{\rho }-it^{\mu \nu \lambda \rho }j_{\rho }
\end{array}
\end{equation}
In the `triality' formulation $(k^{\mu },j^{\nu },u^{\beta
},v^{\alpha})$ are considered as neutral elements (the name
``neutral elements" is due to Chevalley$^{\lbrack 8\rbrack }$),
and existence of such neutral elements in Dirac theory is verified
by the special case Eq.\ (27) below.

We are now in the position to study vector representation of Dirac
spinors in 4-dimensional Minkowski space. The vector
representation of a spinor $\Psi =\Psi _{(1)}+\Psi _{(2)}$
(referred to the unit-basis defined above) is realized in the
following way
\begin{equation} %(20)
\begin{array}{ll}
B^{\mu } &\stackrel{\mbox{\scriptsize def.}}{=}
\frac{1}{2}(\bar{v}i\gamma ^{\mu } \Psi -\bar{\Psi }i\gamma ^{\mu
}v) =\frac{1}{2}(\bar{v}i\gamma ^{\mu } \Psi_{(1)} -\bar{\Psi
}_{(1)}i\gamma ^{\mu }v) \\\noa6 N^{\mu } &
\stackrel{\mbox{\scriptsize def.}}{=} \frac{1}{2}(\bar{\Psi }i
\gamma ^{\mu }u-\bar{u}i\gamma ^{\mu }\Psi )
=\frac{1}{2}(\bar{\Psi }_{(2)} i \gamma ^{\mu }u-\bar{u}i\gamma
^{\mu }\Psi_{(2)} ) \\\noa6 \Psi _{(1)} & =B^{\mu }\eta _{\mu \nu
}i\gamma ^{\nu }v \\\noa6 \Psi _{(2)} & =N^{\mu }\eta _{\mu \nu
}i\gamma ^{\nu }u
\end{array}
\end{equation}
Here $B^{\mu }$ and $N^{\mu }$ are  real vectors, with
4-components. They define the vector representation of the
half-spinors $\Psi _{(1)}$ and $\Psi _{(2)}$ respectively. We see
that
\begin{equation} %(21)
\bar{\Psi }\Psi =\bar{\Psi }_{(1)}\Psi
_{(1)}+\bar{\Psi }_{(2)}\Psi_{(2)}=-B^{\mu }\eta _{\mu \nu
}B^{\nu }+N^{\mu }\eta _{\mu \nu }N^{\nu }
\end{equation}
here $\bar{\Psi }$ represents Dirac conjugate of $\Psi$.

The symbol $\check{c}^{\mu \nu \lambda }$ (see Eq.\ (9)) which
characterizes the algebra of biquaternion (and plays crucial role
in our theory) can be expressed in terms of the unit spinors
$u^{\alpha }$ , $v^{\alpha }$ and Dirac's gama matrix
\begin{equation} %(20)
\check{c}^{\mu \nu \lambda }=\frac{i}{2}\bar{v} (\gamma ^{\mu
}\gamma ^{\nu }\gamma ^{\lambda }+ \gamma ^{\lambda }\gamma ^{\nu
}\gamma ^{\mu })u -\frac{1}{2}\bar{u}(\gamma ^{\mu }\gamma ^{\nu
}\gamma ^{\lambda }- \gamma ^{\lambda }\gamma ^{\nu }\gamma ^{\mu
})u
\end{equation}
Formally, above expression can be considered as the mapping from
$\gamma^\nu \longrightarrow \check{c}^{\mu \nu \lambda }$, or as
the vector representation of Dirac's gamma matrix.

Sometime it is convenient to pass from the unit-basis $(k^{\mu
},j^{\nu },u^{\beta },v^{\alpha })$ to the null-basis $(k_{+}^{\mu
},k_{-}^{\nu },r^{\alpha },l^{\beta })$, where the normalized
``right-handed'' spinor  $r$ and ``left-handed'' spinor  $l$ (pure
spinors) are determined as
\begin{equation} %(23)
\begin{array}{lllll}
u=\frac{1}{2}(r+l) & , & v=\frac{i}{2}(r-l) & , & \bar{r}l=\bar{l}
r=2 \\\noa6 k_{\pm }^{\mu }=\frac{1}{2}(k^{\mu }\pm j^{\mu }) & ,
& k_{\pm }^{\mu }\eta _{\mu \nu }k_{\pm }^{\nu }=0 & , & k_{\pm
}^{\mu }\eta _{\mu \nu }k_{\mp }^{\nu }=\frac{1}{2}
\end{array}
\end{equation}
It is easy to prove that
\begin{equation} %(24)
\begin{array}{ccc}
k_{\mu }\gamma ^{\mu }r=k_{\mu }^{-}\gamma ^{\mu }r=l & , & k_{\mu
}^{+}\gamma ^{\mu }r=0 \\\noa3
k_{\mu }\gamma ^{\mu }l=k_{\mu }^{+}\gamma ^{\mu }l=r & , & k_{\mu
}^{-}\gamma ^{\mu }l=0
\end{array}
\end{equation}
Thus, expressions (20) can be rewritten in the more useful forms
\begin{equation} %(25)
\begin{array}{rl}
G^{\mu } \stackrel{\mbox{\scriptsize def.}}{=} B^{\mu }+iN^{\mu
}&=\frac{1}{2} (\bar{r}\gamma ^{\mu }R -\bar{L}\gamma^{\mu }l) \\
\noa6 R& = \frac{1}{2}G_{\mu} \gamma^{\mu } l \\
\noa6 L& = \frac{-1}{2}G^{\ast}_{ \mu }\gamma^{\mu } r
\end{array}
\end{equation}
here $R$ and $L$ are Dirac's ``right-handed'' and a
``left-handed'' spinors respectively, $\Psi =R+L$. In this
notations
\begin{equation} %(26)
\begin{array}{rl}
\bar{\Psi }\Psi &=-\frac{1 }{2}(G_{\nu }G^{\nu
}+G_{\nu }^{\ast }G^{\ast \nu })\neq G_{\nu }^{\ast }G^{\nu } \\
\bar{\Psi }\gamma_5 \Psi &= -\frac{1 }{2}(G_{\nu
}G^{\nu}-G_{\nu }^{\ast }G^{\ast \nu }) \\
\noa6 \bar{\Psi }\gamma^{\mu}\Psi &=G_{\nu }^{\ast }{c}^{\nu \mu
\lambda }G_{\lambda } \\\noa4 \bar{\Psi }\gamma^{5}
\gamma_{\mu}\Psi &=G_{\nu }^{\ast }\check{c}^{\nu \mu \lambda
}G_{\lambda }
\end{array}
\end{equation}

In order to understand our idea easily, it is convenient to work
in a special coordinate system such that
\begin{equation} %(27)
\begin{array}{ll}
\lbrack u^{\alpha }\rbrack ^{T} & =\lbrack 1,0,0,0\rbrack \\\noa2
\lbrack v^{\alpha }\rbrack ^{T} & =\lbrack 0,0,i,0\rbrack \\\noa2
k^{\mu } & =(1,0,0,0) \\\noa2 j^{\mu } & =(0,0,0,1)
\end{array}
\end{equation}
(here we use the Dirac representation of gamma matrices). In the
physical literature $k^\mu$ means a ``time-direction" and $j^\mu$
is considered as a ``z-direction", which is needed in the study
spin-structure of physical particle. In this special coordinate
system
\begin{equation} %(28)
\Psi^{\alpha } =\Psi^{\alpha } _{(1)} +\Psi^{\alpha }
_{(2)}=\left(
\begin{array}{c}
B^{3} \\
B^{1}+iB^{2} \\
B^{0} \\
0
\end{array}
\right)
 +\left(
\begin{array}{c}
iN^{0} \\
0 \\
iN^{3} \\
-N^{2}+iN^{1}
\end{array}
\right)
\end{equation}

Now let $A^{\mu }\in M^{1+3}$ be a real vector in Minkowski space,
$\Psi _{(1)}\in S_{1}$ and $\Psi _{(2)}\in S_{2}$ are two
half-spinors referred to the unit-basis defined above. The special
cubic-form is defined as
\begin{equation} %(29)
A_{\mu }(\bar{\Psi }_{(1)}\gamma ^{\mu }\Psi _{(2)}+%
\bar{\Psi }_{(2)}\gamma ^{\mu }\Psi _{(1)})=2A_{\mu }(\epsilon
^{\nu \mu \lambda \sigma }k_{\sigma })B_{\nu }N_{\lambda }
\end{equation}
One realizes that there exists an automorphism $\Sigma _{3}$ of order 3 in $%
M^{1+3}\times S_{1}\times S_{2}$, which leaves the trilinear-form
(29) invariant. $\Sigma _{3}$ maps $M$
onto $S_{1}$, $S_{1}$ onto $S_{2}$, and $S_{2}$ onto $M$. (Or equivalently $%
A\rightarrow B \rightarrow N \rightarrow A$.) This result can be
considered as generalization of Cartan's principle of triality$^{[4,8]}$.

Remarks:

(1) More general cubic-form is
\begin{equation} %(30)
C=A_{\mu}\bar{\Psi}\gamma^{\mu}\Psi =A_{\mu} t^{\nu \mu
\lambda }(B_{\nu}B_{\lambda}+N_{\nu}N_{\lambda}) +2A_{\mu}
\epsilon^{\nu \mu \lambda }B_{\nu}N_{\lambda}\,.
\end{equation}
Let operators $q_1$ and $q_2$ are elements of Artin braid group
\begin{equation} %(31)
\begin{array}{lll}
q_1 \Phi(A,B,N)\stackrel{\mbox{\scriptsize def.}}{=}\Phi(-B,A,N) &
, & q_{1}^{-1} \Phi(A,B,N)\stackrel{\mbox{\scriptsize
def.}}{=}\Phi(B,-A,N)
\\\noa3 q_2 \Phi(A,B,N)\stackrel{\mbox{\scriptsize def.}}{=}\Phi(A,-N,B)
& , & q_{2}^{-1} \Phi(A,B,N)\stackrel{\mbox{\scriptsize
def.}}{=}\Phi(A,N,-B)\,.
\end{array}
\end{equation}
In general $q_1 q_1 \neq I$ (is not identical),
$(q_1)^4=(q_2)^4=I$ and $q_1 q_2 q_1 = q_2 q_1 q_2$. We can prove
that for cubic-form (30)
\begin{equation} %(32)
q_1 q_2 q_2 q_1 C(A,B,N)=C(A,B,N)
\end{equation}
That is so-called ``Dirac's game" for spinor rotations.

(2) Let $A^{\mu }$, $B^{\nu }$, $C^{\lambda }$, $D^{\rho}$ are
vectors. We can prove that under permutations
\begin{equation} %(33)
\begin{array}{llr}
t_{\mu \nu \lambda \rho } A^{\mu}B^{\nu }C^{\lambda }D^{\rho }&=&
 t_{\mu \nu \lambda \rho } B^{\mu}C^{\nu
}D^{\lambda }A^{\rho } \\\noa3
\epsilon_{\mu \nu \lambda \rho }A^{\mu}B^{\nu }C^{\lambda }D^{\rho
}&=&-\epsilon_{\mu \nu \lambda \rho } B^{\mu}C^{\nu }D^{\lambda
}A^{\rho }
\end{array}
\end{equation}

\section{Vector representation of Dirac equation.}

The most interesting for physicists is: by passing from ordinary
spinor representation to the vector representation, one can
express Dirac Lagrangian in the Bosonic form
\begin{equation} %(34)
\begin{array}{ll}
&\frac{1}{2}\lbrack \bar{\Psi }i\gamma ^{\mu }(\partial _{\mu
}-ieA_{\mu })\Psi -((\partial _{\mu }+ieA_{\mu })\bar{\Psi
})i\gamma ^{\mu }\Psi \rbrack -m\bar{\Psi }\Psi
\\\noa6 = &\frac{1}{2}\lbrack (\stackrel{e}{\nabla
}_{\mu }G_{\nu })^{\ast }i\check{c}^{\nu \mu \lambda }G_{\lambda
}-G_{\nu }^{\ast }i\check{c}^{\nu \mu \lambda }\stackrel{e}{\nabla
} _{\mu }G_{\lambda }+m(G_{\nu }^{\ast }G^{\ast \nu }+G_{\nu
}G^{\nu })\rbrack
\end{array}
\end{equation}
here
\begin{equation} \stackrel{e}{\nabla }_{\mu }G_{\lambda}
\stackrel{\mbox{\scriptsize def.}}{=}\partial _{\mu }G_{\lambda
}-ieA_{\mu }\eta _{\lambda \rho }c_{5}^{\rho \sigma }G_{\sigma }
\end{equation}
and $A_{\mu }$ is the electromagnetic field. The symbol
$\check{c}^{\nu \mu \lambda}=\check{t}^{\nu \mu
\lambda}-i\check{\epsilon}^{\nu \mu \lambda}$ characterize the
algebra of biquaternion, it contains antisymmetric \textit{Lie}
and symmetric \textit{Jourdan} structure (see Eqs.(13)(14)).

The corresponding massive Dirac equation in the
vector-representation is
\begin{equation} %(35)
i\check{c}^{\mu \nu \lambda } \stackrel{e}{\nabla }_{\nu }
G_{\lambda }-mG^{\mu \ast }= 0{.}
\end{equation}
Or equivalently, takes following self-dual form
\begin{equation}
\begin{array}{l}
G_{\mu \nu }=\frac{i}{2}\ \epsilon _{\mu \nu \lambda \rho
}G^{\lambda \rho }\\
\partial _{\mu }G^{\mu }-imj_{\mu }G^{\mu \ast }=0
\end{array}
\end{equation}
here for simplicity we take $A_{\mu }=0$, and
\begin{equation}
G_{\mu \nu }\stackrel{def.}{=}\partial _{\mu }G_{\nu }-\partial
_{\nu }G_{\mu }+imj_{\mu }G_{\nu }^{\ast }-imj_{\nu }G_{\mu
}^{\ast}
\end{equation}

Sometimes it is convenient to work with `chiral' biquaternions.
Let
\begin{equation} %(39)
\begin{array}{l}
\mathcal{R}^{\mu} \stackrel{\mbox{\scriptsize
def.}}{=}\frac{1}{2}(\eta^{\mu \nu}+c^{\mu \nu}_5 )G_{\nu}
\\\noa6 \mathcal{L}^{\mu} \stackrel{\mbox{\scriptsize
def.}}{=} \frac{1}{2}(\eta^{\mu \nu}-c^{\mu \nu}_5 )G_{\nu}
\end{array}
\end{equation}
Then the equation of motion eq.(36) can be rewritten in the
following form
\begin{equation} %(40)
\begin{array}{l}
\check{c}^{\mu \nu \lambda }i(\partial _{\nu }-ieA_{\nu
})\mathcal{R}^{\lambda}-m\mathcal{L}^{\ast \mu}=0 \\\noa3
\check{c}^{\mu \nu \lambda }i(\partial _{\nu }+ieA_{\nu
})\mathcal{L}^{\lambda}-m\mathcal{R}^{\ast \mu}=0
\end{array}
\end{equation}

The Lagrangian (34) is invariant under $U(1)$ gauge
transformations
\begin{equation} %(41)
\begin{array}{lll}
\tilde{\Psi }= e^{i\alpha }\Psi=(\cos \alpha +i\sin \alpha
)\Psi & , & \tilde{A}_{\mu}=A_{\mu }+\partial _{\mu }\alpha
\end{array}
\end{equation}
In the vector representation it is equivalent to
\begin{equation} %(42)
\tilde{G}^{\mu}=(\eta ^{\mu \nu }\cos \alpha +ic_{5}^{\mu \nu }\sin
\alpha )G_{\nu }=(\exp i\alpha c_{5})^{\mu \nu }G_{\nu }
\end{equation}
It looks like a chiral transformation for $G^{\mu }$, here $(c_{5}^{\mu
\sigma }\eta _{\sigma \rho }c_{5}^{\rho \nu })=\eta ^{\mu \nu }$, and from
which the De Moivre theorem is deduced $(\cos \alpha +ic_{5}\sin \alpha
)^{n}=(\cos n\alpha +ic_{5}\sin n\alpha )$.

We know that the massless Dirac Lagrangian is invariant under the chiral
transformation
\begin{equation} %(43)
\begin{array}{lll}
\Psi \rightarrow e^{ia\gamma _{5}}\Psi & ; & \bar{\Psi }\rightarrow
\bar{\Psi }e^{ia\gamma _{5}}
\end{array}
\end{equation}
In the vector representation it is equivalent to
\begin{equation} %(44)
G_{\mu }\rightarrow e^{ia}G_{\mu }
\end{equation}
It means that a chiral transformation for $\Psi $ is equivalent to
a $U(1)$ transformation for $G_{\mu }$. Therefore we must
distinguish between the plane wave solutions for $\Psi$ and for
$G^{\mu }$.

The relationship between the \textit{first} order Dirac equation
and the \textit{first} order Maxwell equation  is a subject of
interest of investigators since the time of creation of quantum
mechanics. We will prove that only 3D-vector part of biquaternion
has analogy with Maxwell's field strength $\vec{E}+i\vec{H}$.

In electromagnetic theory it is convenient to use the real
antisymmetric field-strength tensor $F_{[\mu \nu]}$ instead of
$\vec{D}=\vec{E}+i\vec{H}$. The antisymmetric property of $F_{[\mu
\nu]}$ implies that $F_{[\mu \nu]}$ including only six independent
real components and they have one to one correspondence with
$\vec{E}$ and $\vec{H}$, such that (real constant)
$Re(\vec{D}\cdot\vec{D})=(\vec{E}\cdot\vec{E}-\vec{H}\cdot\vec{H})=\frac{-1}{2}F_{[\mu
\nu]}F^{[\mu \nu]}$.

Similarly, in our theory we introduce complex tensor $H_{\mu \nu}$
which satisfies
\begin{equation} %(45)
\left(\frac{-1}{2}{c}^{\mu \lambda \nu \rho}\right)H_{\lambda \rho}= H^{\mu
\nu}
\end{equation}
here ${c}^{\mu \nu \lambda \rho}$ is defined by (5). The above
algebraic equation can be considered as a ``\textbf{self duality}"
condition for complex tensor $H_{\mu \nu}$ because of
\begin{equation} %(46)
\left(\frac{-1}{2}{c}^{\mu \lambda \nu
\rho}\right)\left(\frac{-1}{2} {c}^{\lambda \sigma \rho
\delta}\right)={g}^{\mu \sigma} {g}^{\nu \delta}
\end{equation}
The ``self duality" condition implies that $H_{\mu \nu}$ includes
only \textit{four} independent \textit{complex} components and
there is one to one correspondence with complex $G^\mu$. That is
\begin{equation} %(47)
\begin{array}{rcl}
H^{\ast \nu \lambda}&\stackrel{\mbox{\scriptsize def.}}{=}&
\check{c}^{\lambda \nu \mu} G_{\mu} \\\noa3 H^{\nu \lambda}&=&
G^{\ast}_{\mu} \check{c}^{\mu \nu \lambda}
\\\noa6 G^{\ast \mu}&=&\frac{1}{2} \check{c}^{\nu \mu \lambda}
H_{\nu \lambda}=- H^{\mu \nu}j_\nu
\\\noa6 G^{\mu}&=&\frac{1}{2} H^{\ast}_{ \nu
\lambda} \check{c}^{\lambda \mu \nu} =-H^{\ast \mu \nu}j_\nu
\end{array}
\end{equation}
and $ (G_{\mu}G^{\mu})^\ast=\frac{1}{4}(H_{\mu\nu}H^{\mu\nu})$.
Notice that the self-dual tensor $H_{\mu \nu}$ containing both
(antisymmetric) Lie and (symmetric) \textit{Jourdan} structure. In
these notations the (uncharged) Dirac equation (36) can be
rewritten in the following form
\begin{equation} %(48)
\partial_{\nu}H^{\nu \mu}-\frac{im}{2}
(\check{c}^{\nu \mu \lambda}H_{\nu \lambda})^{\ast}=0
\end{equation}

We know that $G_{\mu }$ can be decomposed into the sum of two
parts (see Eq.\ (12)):
\begin{equation} %(49)
G^{\mu }=p_{+}^{\mu \nu }G_{\nu}+p_{-}^{\mu \nu }G_{\nu}
\mathcal{G}^{\mu}+(-j^{\mu}\phi)
\end{equation}
Here $\mathcal{G}^{\mu}$ is the 3d-vector part of biquaternion and
$\phi=\phi_{1}+i\phi_{2}$ is the (complex) scalar. Corresponding
to them, $H_{\mu \nu}$ can be decomposed into the sum of symmetric
and antisymmetric parts $H_{\mu \nu}=H_{(\mu \nu)}+H_{[\mu \nu]}$.
Self-duality condition (45) implies that
\begin{equation} %(50)
\begin{array}{llll}
H^{(\mu \nu)}&= (\frac{1}{4}H^{(\lambda \rho)} \eta_{\lambda
\rho})\eta^{\mu \nu}&=& {\phi}^{\ast}\eta^{\mu \nu} \\\noa6
H^{[\mu \nu]}&=\frac{-i}{2}{\epsilon}^{\mu \nu \lambda \rho}
H_{[\lambda \rho]}&=&
\mathcal{G}^{\ast}_{\lambda}\check{c}^{\lambda \mu \nu}
\end{array}
\end{equation}
The last relation means that 3d complex vector
$\mathcal{G}_{\lambda}$ corresponds to anti-self-dual
antisymmetric $H^{[\mu \nu]}$ which includes \textit{three}
independent \textit{complex} components. Without loss of
generality, anti-self-dual $H_{[\mu \nu]}$ can be expressed in
terms of real antisymmetric $h^{[\mu \nu]}$ or $F_{[\mu \nu]}$ in
the following way
\begin{equation} %(51)
\begin{array}{rl}
H^{[\mu \nu]}& =h^{[\mu \nu]}-\frac{i}{2} \epsilon ^{\mu \nu
\sigma \rho}h_{[\sigma \rho]} \\\noa6 h^{[\mu \nu]}&
=-\frac{1}{2}\left(F^{[\mu \nu]}+\frac{1}{2} \epsilon ^{\mu \nu
\sigma \rho}{F}_{[\sigma \rho]}\right)
\end{array}
\end{equation}
In this notation, uncharged Dirac equation takes the compact form
\begin{equation} %(52)
\begin{array}{rrl}
\stackrel{-m}{\nabla }_{\nu }{F}_{[\mu \nu]}&  = &
-\stackrel{m}{\nabla }_{\mu } (\phi_{1}-\phi_{2}) \\\noa6
\frac{1}{2}\epsilon ^{\mu \nu \lambda \rho}\stackrel{m}{\nabla
}_{\nu } {F}_{[\lambda \rho]}& = & -\stackrel{-m}{\nabla }_{\mu
}(\phi_{1}+\phi_{2})
\end{array}
\end{equation}
here $\stackrel{m} {\nabla }_{\nu }=(\partial_{\nu}+mj_{\nu})$ and
$\stackrel{-m} {\nabla }_{\nu }\stackrel{m} {{\nabla }^{\nu
}}=(\partial\partial+m^2 )$ . It is interesting to notice the
similarity between these equations and the Maxwell equations which
include ``electric" and ``magnetic" currents (and making no
reference to the vector potential).

The most interesting special case is to take
$(\phi_{1}+\phi_{2})=0$. In this special case the last equation in
(52) looks like Bianchi identity.

In fact, Dirac equation (36) and Dirac Lagrangian (34) are
invariant under U(1) gauge transformations (41) and (42), where
$\alpha (x)$ depends on spacetime position. All dependence on
$\alpha (x)$ drops out of Lagrangian, so that we can simply
replace $G^{\mu}$ and $A_{\mu}$ everywhere with $\tilde{G}^{\mu}$
and $\tilde{A}_{\mu}$. (In the vacuum, i.e., $\vec{E}=\vec{H}=0$,
a corresponding potential $A_{\mu}=\partial_{\mu}\theta$ is a pure
gauge, here $\theta (x)$ is \textit{arbitrary} function of $x$).
This is a choice of gauge, fixed by imposing the special condition
on $\tilde{G}^{\mu}$ rather than by imposing condition on the
gauge field themselves.  Specially, it is possible choice the
gauge such that $(\tilde{\phi}_{1}+\tilde{\phi}_{2})=0$
(spontaneous symmetry breaking). In this gauge (and in the absence
of electromagnetic potential, i.e.,
$\tilde{A}_{\mu}=\partial_{\mu}\tilde{\theta}=0$) the second
equation in Eq.\ (52) can be considered as a Bianchi identity. And
it is precisely the condition enabling one to write
\begin{equation} %(53)
{F}_{[\mu \nu]}=\partial_{\mu}B_{\nu}-\partial_{\nu}B_{\mu}
+mj_{\mu}B_{\nu}-mj_{\nu}B_{\mu}
\end{equation}
here $B_{\mu}$ is a gauge potential (real), and the field strength
${F}_{[\mu \nu]}$ is invariant under the following ``gauge
translations"
\begin{equation} %(54)
\tilde{B}_{\mu}=B_{\mu}+\partial_{\mu}\beta+mj_{\mu}\beta
\end{equation}
here $\beta (x)$ is a gauge freedom.

\section{``Parent action'' approach}

\vspace{1pt}The ``parent (or master) action'' approach was
proposed by Deser and Jackiw$^{\lbrack 2\rbrack }$ to establish,
\textit{at the level of the Lagrangian} instead of equation of
motion, the equivalence (by a Legender transformation) or the
so-called duality between the first order 3D massive self-dual
vector field and two order 3D topologically massive gauge
field$^{\lbrack 3\rbrack }$. We propose here an alternative way to
generalize it to Dirac Lagrangian. The reason why such cases are
important and interesting is the fact that duality typically
exchanges the coupling regimes $g\longrightarrow 1/g$, then the
weak coupling regime in one model is the strong regime in the
other and {\it vice versa}. Knowing the explicit relation thus
allows perturbative calculations in the variables of the original
theory both in the strong and weak coupling  regimes. Moreover,
the above approach gave us the possibility to define the
\textit{potential} (gauge) field to corresponding vector field and
making no reference to the so-called ``Bianchi identity".

Let $G_{\mu }$ is the Dirac field and its dual (potential) field
is denoted by $H_{\mu }$. Consider the so-called  parent
action $\int (d^{4}x)S_{GH}$
\begin{equation} %(55)
\begin{array}{lrl}
S_{GH}= &\left(G_{\nu} +\frac{1}{2}H_{\nu
}\right)^{\ast}i\check{c}^{\nu \mu
\lambda}\left(\stackrel{ea}{\nabla }_{\mu }H_{\lambda}\right)
&+\frac{m-a}{2}G_{\nu }G^{\nu } \\\noa6 & -\left(
\stackrel{ea}{\nabla }_{\mu }H_{\nu }\right)^{\ast }
i\check{c}^{\nu \mu \lambda }\left(G_{\lambda}
+\frac{1}{2}H_{\lambda }\right)& + \frac{m-a}{2}G_{\nu}^{\ast
}G^{\ast \nu }+\partial _{\mu }(Z_{0}^{\mu})
\end{array}
\end{equation}
here $ \stackrel{ea}{\nabla }_{\mu }H_{\lambda}
\stackrel{\mbox{\scriptsize def.}}{=}(\partial _{\mu }H_{\lambda
}-ieA_{\mu }\eta _{\lambda \rho }c_{5}^{\rho \sigma }H_{\sigma
}-iaj_\lambda H^{\ast}_\mu)$ and ``$a$" is a real constant ($a
\neq m$). Varying $S_{GH}$ with respect to $H_{\mu }$, gives
directly
\begin{equation} %(56)
\check{c}^{\nu \mu \lambda }\stackrel{ea}{\nabla }_{\mu }
G_{\lambda }=-\check{c}^{\nu \mu \lambda } \stackrel{ea}{\nabla
}_{\mu }H_{\lambda }
\end{equation}
Plugging this back into $S_{GH}$, eliminating $H_{\mu }$ from it,
gives Dirac Lagrangian (34) which is linear in derivatives
\begin{equation} %(57)
\begin{array}{ll}
S_{G}= &\frac{1}{2}\lbrack (\stackrel{e}{\nabla }_{\mu }G_{\nu
})^{\ast }i\check{c}^{\nu \mu \lambda }G_{\lambda }-G_{\nu }^{\ast
}i\check{c}^{\nu \mu \lambda }\stackrel{e}{\nabla } _{\mu
}G_{\lambda }+m(G_{\nu }^{\ast }G^{\ast \nu }+G_{\nu }G^{\nu
})\rbrack \\\noa6 & +\partial _{\mu }\lbrack Z_{0}^\mu-
\frac{1}{2}(H_{\nu }^{\ast }i\check{c}^{\nu \mu \lambda
}G_{\lambda }-G_{\nu }^{\ast }i\check{c}^{\nu \mu \lambda
}H_{\lambda })\rbrack
\end{array}
\end{equation}
The additional last term is the total divergence and does not
contribute to equation of motion. We choice $Z_{0}^\mu$ such that
the last surface term is equal to zero. We want to notice that
elimination of $H^{\mu }$ from $S_{GH}$ is not so easy here;
the solution of equation (56) for $H^{\mu }$ is non-trivial.
However, we have not attempted to an explicit solution of Eq.\ (56)
here.

Alternatively we may first vary $G_{\nu }$ in the parent action $\int
(d^{4}x)S_{GH}$, which yields
\begin{equation} %(58)
(m-a)G^{\lambda }=( \stackrel{ea}{\nabla }_{\mu } H_{\nu })^{\ast
} i\check{c}^{\nu \mu \lambda }
\end{equation}
(In fact, this expression can be deduced from (36) and (56)).
Plugging this back into $S_{GH}$, eliminating $G_{\nu }$ from it,
we obtain
\begin{equation} %(59)
\begin{array}{ll}
S_{H}= &\frac{1}{2(m-a)} (\check{c}^{\nu \mu \lambda }
\stackrel{ea}{\nabla }_{\mu }H_{\lambda }) (\check{c}^{\nu \rho
\delta }\stackrel{ea}{\nabla }_{\rho}H_{\delta}) +\frac{1}{2}
H_{\nu }^{\ast }i\check{c}^{\nu \mu \lambda} (\stackrel{ea}{\nabla
}_{\mu }H_{\lambda }) \\\noa6 &+[\cdots]^{\ast}
+\partial_\mu(Z_{0}^\mu)
\end{array}
\end{equation}
which is quadratic in derivative. Here the term $[\cdots]^{\ast }$ denotes complex
conjugated part of the former
parts. Thus, from the parent action $\int (d^{4}x)S_{GH}$ we have
shown that the Dirac action $\int (d^{4}x)S_{G}$ (i.e.\ (57)) and new
action $\int (d^{4}x)S_{H}$ (i.e.\ (59)) is dual to each other; the
two actions represent the same physics (at least classically), but
the physical description is given using different fields.

The corresponding equation of motion for new field $H^{\mu }$ is
\begin{equation} %(60)
\partial\partial H^{\lambda }-
(m+a)\partial_{\mu }H_{\nu }^{\ast }
i\check{c}^{\nu \mu \lambda }-maH^{\lambda }=0
\end{equation}
(here for simplicity we take $A_{\mu}=0$.) Clearly it is identical
with Dirac equation (36) when equation (58) is used.

If $ a+m=0 $ then equation (60) is nothing but the Klein--Gordon
equation. But here we are interested only in the following $ a=e=0 $
case, i.e.
\begin{equation} %(61)
S_{H}=  \frac{-b^2}{2m}\lbrack
\partial_{\mu }H_{\nu } \partial^{\mu}H^{\nu}
-m H_{\nu }^{\ast }i\check{c}^{\nu \mu \lambda}
\partial_{\mu }H_{\lambda }-\partial_\mu (H^{\mu} \partial H)\rbrack
+\lbrack \cdots\rbrack ^{\ast }
\end{equation}
\begin{equation} %(62)
\partial\partial H^{\lambda }-
m\partial_{\mu }H_{\nu }^{\ast } i\check{c}^{\nu \mu \lambda }=0
\end{equation}
Here we have changed the notation from (59), with the replacement
$H^{\mu }\rightarrow bH^{\mu }$  ($b$ is dimensional real
constant).  The novel mass term in (61) is the cubic-form
(three-form) and bilinear in the mass $m$ and in the derivative.
This kind of a mass term one can find in the theory of
topologically massive gauge field, and it was analyzed completely
in Ref.\ [3]. At the quantum level, it suggests a new cure for the
infrared problem, without disturbing the ultraviolet or gauge
aspects. The equation of motion (62) for new field $H^{\mu }$ is
identical with (uncharged) Dirac equation (36) when
\begin{equation} %(63)
mG^{\lambda }=( \partial_{\mu } H_{\nu }^{\ast }) i\check{c}^{\nu
\mu \lambda }
\end{equation}
is used.

In classical dynamics, the force $\vec{F}$ appears explicitly in
the equation of motion of particle $m\vec{a}=\vec{F}$. It is true
that a potential $V$ can be introduced $\vec{F}=-{\nabla} V$, but
all potentials differing from $V$ by a \textit{constant} give the
same $\vec{F}$, so that $V$ is not uniquely determined by
$\vec{F}$. Similarly, in classical electromagnetic theory the
basic laws (Maxwell's equation, together with Lorentz's force
expression) are all explicitly expressed in terms of the fields
strength $\textbf{E}$ and $\textbf{H}$. One may introduce the
potentials $(\textbf{A},\phi)$ such that $\textbf{E}=-\nabla
\phi-\frac{1}{c} \frac{\partial {\textbf{A}}}{\partial {t}}$ and
$\textbf{H}=-\nabla \times \textbf{A}$. But under the gauge
transformation to new potentials $\tilde{\textbf{A}}=
\textbf{A}+\nabla \chi$ and $\tilde{\phi}=\phi-\frac{1}{c}
\frac{\partial\chi}{\partial t}$ where $\partial\partial\chi=0$,
the fields $\textbf{E}$ and $\textbf{H}$ and Maxwell equation
together with the Lorentz gauge condition remain invariant. The
following theorem due to Bateman (1904): the \textit{general}
real-analytic solution of the wave equation
$\partial\partial\chi=0$ in flat space-time is
\begin{equation} %(64)
\chi=\int^\pi_{-\pi}\Theta(\zeta ,\xi ,\theta)d\theta
\end{equation}
here
\begin{equation} %(65)
\begin{array}{ll}
\zeta=(v^1_{\mu} \cos\theta+v^2_{\mu} \sin\theta+iv^3_{\mu})
x^{\mu}=\zeta_\mu x^{\mu}\\\noa3 \xi=(v^0_{\mu}
\cos\theta+v^2_{\mu}+i v^3_{\mu} \sin\theta) x^{\mu}=\xi_\mu
x^{\mu}
\end{array}
\end{equation}
and  $v^a_\mu$ is  vierbein, $v_{\mu}^a \eta^{\mu \nu}v_{\nu}^b
=\eta^{ab}$, without loss of generality we take
$v^a_\mu=\delta^a_\mu$. $\Theta(\zeta ,\xi ,\theta)$ is an
\textit{arbitrary} function of \textit{three} variables,
complex-analytic in the first two$^{[15][16]}$. In fact the
function $\Theta$ is, in effect, a function on (projective)
twistor space, and twistor theory provides a way of understanding
and generalizing this solution formula$^{[16]}$. We find that the
field equation for ``gauge freedom" $\partial\partial
\chi(x,y,z,t)=0$ has \textit{disappeared} in passing to ``Bateman
function" $\Theta(\zeta ,\xi ,\theta)$ which does not subjected to
any different equation.

In our theory, $G^{\mu}$ appears explicitly in the equation of
motion of Dirac's particle. Thus $G^{\mu}$ can be considered as
the field strength while $H_{\nu}$ in expression (63) can be
considered as the corresponding potential. Potential is not
uniquely determined by field strength. In general, let ${H}^{\mu
}\rightarrow H^{\mu }+ \delta H^{\mu}$ denotes the `gauge
transformation", then only when $\delta H^{\mu}$ satisfies
\begin{equation} %(66)
\check{c}^{\nu \mu \lambda } \partial_{\mu }(\delta H_{\lambda
})=0
\end{equation}
the expression (63) and the equation of motion Eq.\ (62) are
unchanged. Notice, the mass term in Lagrangian (61) is changed
under the above transformations! However it changes only by a
total derivative. Modification of a Lagrangian by a total
derivative does not affect equations of motion; that is why
equations of motion are ``gauge" covariant even though $S_H$ is
not ``gauge" invariant. And this implies the topological property
of the mass term. Moreover the equation of motion (62) yields
\begin{equation} %(67)
\partial_\mu J^\mu =\partial_{\mu}[i H_{\nu}c^{\nu \lambda}_5
\partial_\mu H_{\lambda}-i(H_{\nu}c^{\nu \lambda}_5
\partial_\mu H_{\lambda})^\ast
+m H^{\ast }_{\nu}c^{\nu \mu \lambda}H_{\lambda}]=0
\end{equation}
In fact, the above $J^\mu$ can be considered as the Noether's
current, it is not invariant object, but it is conserved in the
ordinary sense.

The general solution of Eq.\ (66) for $(\delta H_{\mu})$ is
non-trivial, and we have not attempted to an explicit general
solution of Eq.\ (66) here. The fundamental reason for non-trivial
property of ``gauge freedom" is that $\check{c}^{\nu \mu \lambda}$
containing both (antisymmetric) Lie and (symmetric)
\textit{Jourdan} structure (see Eq.\ (14)). In equation
$\check{c}^{\nu \mu \lambda }\partial_{\mu }(\delta H_{\lambda
})=0$ the ``gauge freedom" $\delta H_{\lambda }$ can be considered
as a ``massless particle" in free motion, or as a Goldston
particle. In fact $\delta H_{\lambda }$ is not physical particle,
the above consideration is only for visualization purposes. The
study of ``gauge freedom" is equivalent to study of physically
meaning vacuum, and in our opinion, the physically meaning pure
gauge vacuum must be not to mention the coding of any specific
field equation.

In relativistic field theories one deals with hyperbolic equations
in space-time: for example, the wave equation, the Dirac equation,
Maxwell's equation, the Yang--Mills equation, Einstein's equation,
and so forth. What all these equations have in common is that the
general solution of the equation depends on one or more
\textbf{arbitrary} functions of \textbf{three} variables and which
\textit{do not subject to any different equation}. For example, in
the hyperbolic case, the specification of initial data on a
(three-dimensional) space-like hypersurface uniquely determines a
solution throughout space-time. Thus the above fields may be
regarded as those defined on some \textit{three}-parameter initial
sets (Cauchy hypersurface) and thence extended over the rest of
the space by the field equation. From the physical point of view,
it seems likely that the reason why this work is closely tied up
with the fact that the wave equation in \textit{flat} space-time
satisfies \textbf{Huygens Principle} (HP). (Some possible
formulations of ``Huygens Principle" one can finds in Ref.\ [16].)

Consider the transformation $H_{\mu} \rightarrow H_{\mu}+\delta
H_{\mu}=H_{\mu}+\partial_{\mu}\Theta$. Here
$\Theta(\zeta,\xi,\theta)$ is an arbitrary function of three
variables (defined as in Eqs.\ (64)and (65)), which characterizes the gauge
freedom in the potential $H_{\mu}$ and does \textit{not subject
to any differential equation}. Plugging this into Eq.\ (66), we get
\begin{equation} %(68)
\check{c}^{\nu \mu \lambda } \partial_{\mu }(\delta
H_{\lambda})=\check{c}^{\nu \mu \lambda } \partial_{\mu
}(\partial_{\lambda}\Theta)=j^\nu
\partial
\partial \Theta=0
\end{equation}
and this means that under transformation $H_\mu \rightarrow
H_\mu+\displaystyle\sum_{\theta} \partial_\mu \Theta$ , the
equation of motion Eq.\ (62) is unchanged, while the Lagrangian
(61) is changed only by a total derivative. It is important to
notice that two \textit{null} complex vectors (in Eq.\ (65))
$\zeta_\mu=\partial_{\mu} \zeta$ and $\xi_\mu=\partial_{\nu} \xi$
are orthogonal (i.e.\ $\zeta_\mu \zeta^\mu=\xi_\nu
\xi^\nu=\zeta_\mu \xi^\mu=0$) and $(\zeta^{\mu} \xi^{\nu}
-\xi^{\mu} \zeta^\nu)=\frac{i}{2}\epsilon^{\mu \nu \lambda
\rho}(\zeta_{\lambda} \xi_{\rho} -\xi_{\lambda} \zeta_\rho)$ is a
\textit{self-dual} bivector. This means that the gauge freedom
$\delta H_{\mu}$ is localized only on self-dual totally null
projective $\alpha$-plane in complexified Minkowski space, whose
tangent vectors are $\zeta_{\mu}$ and $\xi_{\mu}$. We find that
complicated differential equation for ``gauge freedom" (i.e.\ Eq.\
(66)) has \textit{disappeared} in passing to generalized ``Bateman
function" on a \textbf{self-dual totally null} projective
$\alpha$-plane.

In general, from mathematical point of view the $\delta H_\mu$ in
Eq.\ (66) is not necessarily $\delta H_\mu=\partial_\mu \Theta$.
In other words, the pure-gauge vacuum is not necessarily
$H_\mu=\partial_\mu \Theta$, because $H_\mu$ is not a connection
of the common curvature tensor. For better understanding the
situation, it is convenient to use self-dual tensor $H_{\mu \nu}$
which was introduced in the previous section and defined by Eqs\
(45), (47), and (63),
\begin{equation} %(69)
mG^{\ast \mu}=\frac{1}{2}\check{c}^{\nu \mu \lambda} H_{\nu
\lambda}=-H^{\mu \nu}j_\nu
\end{equation}
\begin{equation} %(70)
\begin{array}{rcl}
\noa6 H^{\mu \nu}&\stackrel{\mbox{\scriptsize
def.}}{=}&\displaystyle i c^{\mu \lambda \rho
\nu}\partial_{\lambda}H_{\rho}
\\\noa6 &=&\displaystyle i[\eta^{\mu \nu}\partial
H+(\partial^\mu H^\nu -\partial^\nu H^\mu ) \\\noa6 &&
-\frac{i}{2} \epsilon^{\mu \nu \lambda \rho} (\partial_\lambda
H_\rho-\partial_\rho H_\lambda)]
\end{array}
\end{equation}
The self-duality condition (45) implies that $H_{\mu \nu}$
includes only \textit{four} independent \textit{complex}
components and there is one to one correspondence with complex
$G^\mu$ (see Eq.\ (47)). The self-dual tensor $H_{\mu \nu}$ (which
is defined by Eq.\ (70)) can be considered as a ``natural"
generalization of a curvature tensor of a common gauge field. In
the above notations the Lagrangian (61) and the equation of motion
(62) take the following form
\begin{equation} %(71)
S_H=\frac{b^2}{8m}( H_{\mu \nu}H^{\mu \nu}-2mH^{\ast}_{\mu}
\check{c}^{\nu \mu \lambda} H_{\nu \lambda})
-\frac{ib^2}{2m}\partial_{\mu}(H^{\mu \nu}H_\nu)+[\cdots]^{\ast}
\end{equation}
\begin{equation} %(72)
\partial_{\nu}H^{\nu \mu}+im H^{\ast \mu \nu}j_\nu=0
\end{equation}
The above action is dual equivalent to (uncharged) Dirac's action
(34) and the equation of motion (72) is identical with Dirac's
equation (36) when equation (69) is used. Notice again, the novel
self-dual $H^{\mu \nu}$ is not antisymmetric, it contains not only
(antisymmetric) Lie but also (symmetric) \textit{Jourdan}
structures. Thus there is no so-called "Bianchi identity" for
$H^{\mu \nu}$.

The tensor $H^{\mu \nu}$ must be invariant object. The study of
"gauge freedom" is equivalent to study of physically meaning
vacuum. Physical vacuum of Dirac field must be denoted by $G^\mu
=0$ and it is equivalent to $H^{\mu \nu}=0$.  From mathematical
point of view, the pure gauge vacuum $\delta H_\mu$ must satisfies
Eq.\ (66), and it is equivalent to the self-dual equation $c^{\mu
\lambda \rho \nu}\partial_{\lambda}(\delta H_{\rho})=0$ (not
necessary to Eq.\ (68)). This implies the self-dual property of
the pure gauge vacuum.

In my opinion, the self-dual totally null projective
$\alpha$-plane is a good candidate for the physically meaning pure
gauge vacuum. For this
\begin{equation} %(73)
\delta H^\mu=\sum_{\theta}(\zeta_\lambda Q_\nu +\xi_\lambda W_\nu)
\check{c}^{ \nu \lambda \mu}
\end{equation}
where the components of the vectors $Q_\nu(\zeta, \xi, \theta)$
and $W_\nu(\zeta, \xi, \theta)$ are \textbf{arbitrary} functions
of \textbf{three} variables (``Bateman functions") and $\zeta$,
$\xi$ are determined by Eq.\ (65) (in general $\theta$ is not
necessarily real). So again one finds that the different equation
$c^{\mu \lambda \rho \nu}\partial_{\lambda}(\delta H_{\rho})=0$
disappears when one passes to the self-dual totally null (complex)
$\alpha$-plane.

I claim that the pure gauge freedom (73), which we study here
satisfies `\textit{Huygens Principle}', although I shall neither
prove this claim here nor define exactly what I mean by it. In
fact, there is some different formulations of `\textit{Huygens
Principle}'$^{[16]}$ and  up to now there is no exact mathematical
definition for `HP equations'. From physical point of view, they
looks like equation of non-physical `Goldston particles'. The pure
gauge freedom which satisfies `HP equation' must propagates
`cleanly' without back-scattering. Moreover `HP equations' must
have global consequences: essentially they must imply that no
global scattering occurs, that the in-data at past infinity equals
the out-data at future infinity.

Further speculations :  Since K.\ Wilson's seminal work on lattice
gauge theory in 1974, the regularization of quantum field theory
by a space-time lattice has become one of the basic method for
non-perturbative studies in the field theory. The result obtained
with lattice gauge calculations so far substantially improved our
knowledge of QCD. An example of this success can be seen in the
study of quark confinement at strong couplings. Despite these
successes, the fundamental unsolved problems with spinor-fermion
doubling and chiral symmetry, show that we still have an acute
need for new ideas.

The basic reason for fermion doubling is that the common Dirac's
equation for fermion is of \textbf{first order}! However in our
theory after dualization, Dirac's equation become \textbf{two
orders} and as we know there is no ``doubling problem" in two-order
Lagrangian. Secondly, according to Wilson we discretize space into a
simple lattice and naturally associate the vector gauge field with
directed lattice \textbf{link}, antisymmetric tensors with
oriented areas, etc. And geometry scalars and spinors are
associated with lattice \textbf{sites}. In contrast to Wilson's
regularization, in our theory a vector representation of spinor
gives us the possibility to associate vector-fermion with the
lattice \textbf{link}. These ideas have opened up new and exciting
possibilities towards to avoid the problems with lattice fermion.

\section{Topologically massive gauge field}

Field theorists have been greatly enthused over the Chern--Simons
term $I_{CS}$ in resent years. And it is known that only in three
dimensions, a topological term $I_{CS}$, (which must include 3D
Levi--Civita symbol) is bilinear and can produce a mass$^{\lbrack
2\rbrack }$ of the gauge field. The study of such gauge theory has
particular interest because the Chern--Simons characteristics
supply a gauge-invariant mass-generating mechanism. The
Chern--Simons term has been widely used to model various physical
processes in (2+1)-dimensional space-time, that is, phenomena
confined to motion on a plane, like Holl effect.

The above (2+1)-dimensional space-time $M^{2+1}$ (with
Chern--Simons term on it) can be considered as a ``\textit{spin
manifold}", by which we mean a three-manifold with a chosen spin
structure. The three-dimensional Chern--Simons mass term can be
promoting to four-dimensions$^{\lbrack 18\rbrack }$, and this
entails choosing an external fixed embedding 4-dimensional vector.
In our theory $j_{\mu}$ can be considered as such an embedding
vector. On the analogy of decomposition (12) and (49), the
space-time $M^{3+1}$ can be split into two parts, $p_{+}^{\mu \nu
}x_{\nu}\in M^{2+1}$ and $p_{-}^{\mu \nu }x_{\nu}\in M^{1}$, where
the former part is an oriented (2+1)-dimensional ``\textit{spin
manifold}" which is needed here. (Comments: From abstract point of
view the above decomposition depends on the choice of ``neutral
element" $j_{\mu}$. However, in our theory the unit vector $j^{\mu
}$ is offshoot from $\eta_{\mu \nu }$ and $\check{c}^{\nu \mu
\lambda }$. In other words the algebra of biquaternion
characterized by $\eta_{\mu \nu }$ and $\check{c}^{\nu \mu \lambda
}$ only, they uniquely define the unit element $j^{\lambda
}=\frac{1}{4}\eta _{\mu \nu }\check{c}^{\mu \nu \lambda }$.)

Now let us dimensionally reduce dual models discussed in the
previous section. We will apply the dualization rule only to the
vector part of $G$, i.e. $\mathcal{G}^{\mu}=p_{+}^{\mu \nu }G_{\nu
}$. In other words, we consider the duality between
($\mathcal{G}^{\mu}; \phi$) and ($\mathcal{B}^{\mu}; \phi$), where
two models have the same $\phi$. The 3D complex vector
$\mathcal{B}^{\mu }$ (which dual to $\mathcal{G}^{\mu}$) is
defined as $\mathcal{B}^{\mu }\stackrel{\mbox{\scriptsize
def.}}{=}p_{+}^{\mu \nu }B_{\nu }$, where $B_{\nu }$ is a gauge
potential (complex). The corresponding field strength tensor
\begin{equation} %(74)
B_{\mu \nu }\stackrel{\mbox{\scriptsize def.}}{=}\partial_{\mu }B_{\nu
}-\partial_{\nu}B_{\mu}+iaj_{\mu}B_{\nu}^{\ast}-iaj_{\nu}
B_{\mu}^{\ast}
\end{equation}
and ``Bianchi identity" is
\begin{equation} %(75)
\stackrel{a}{\nabla}_{\mu }B_{\nu \lambda} +
\stackrel{a}{\nabla}_{\lambda }B_{\mu \nu} +
\stackrel{a}{\nabla}_{\nu }B_{\lambda \mu}=0
\end{equation}
In this section $\stackrel{a}{\nabla }_{\mu }
\stackrel{\mbox{\scriptsize def.}}{=}(\partial _{\mu }+iaj_{\mu }C)$ here $C$ is the
operator of complex conjugation, $a\neq m$ is a real constant and
$\stackrel{-a}{\nabla }_{\mu } \stackrel{a}{{\nabla }^{\mu
}}=(\partial \partial +a^2)$.

$B_{\mu \nu}$ is unchanged under the following  ``gauge
transformations'' $\tilde{B}_{\mu }=B_{\mu }+\partial_{\mu
}\beta+iaj_{\mu}{\beta}^{\ast} $. We can do gauge transformation
for $B_{\mu }$, such that $\tilde{B}_{\mu }j^{\mu }=0$. In this
``axial gauge" $\tilde{B}_{\mu }\stackrel{\mbox{\scriptsize
def.}}{=}\mathcal{B}_{\mu }$, and the corresponding field strength
(which is needed here)
\begin{equation} %(76)
\mathcal{F}_{\mu \nu }\stackrel{\mbox{\scriptsize def.}}{=}\partial _{\mu
}\mathcal{B}_{\nu }-\partial _{\nu}\mathcal{B}_{\mu }
+iaj_{\mu}\mathcal{B}_{\nu}^{\ast}-iaj_{\nu}
\mathcal{B}_{\mu}^{\ast}
\end{equation}
where $\mathcal{B}_{\mu}j^{\mu}=0$. Up to ``axial gauge"
$\mathcal{F}_{\mu \nu}$ is equivalent to $B_{\mu \nu }$.

The so-called parent action $\int (d^{4}x)S_{\mathcal{G} \phi
\mathcal{B}}$ is
\begin{equation} %(77)
\begin{array}{ll}
S_{\mathcal{G} \phi \mathcal{B}}= &
\frac{1}{2}\left(\mathcal{G}_{\nu }^{\ast
}+\frac{1}{2}\mathcal{B}_{\nu }^{\ast }\right)i\check{c}^{\nu \mu
\lambda }\mathcal{F}_{\mu \lambda }+ \frac{i}{2}(\mathcal{G}^{\mu
}\partial _{\mu }\phi ^{\ast }- \phi ^{\ast }\partial _{\mu
}\mathcal{G}^{\mu}) \\\noa6 &
+\frac{m-a}{2}\mathcal{G}_{\nu}\mathcal{G}^{\nu }+
\frac{i}{2}(\phi ^{\ast }j^{\mu}\stackrel{-m}{\nabla }_{\mu }\phi)
+\lbrack \cdots\rbrack ^{\ast} +\partial_{\mu}(Z^{\mu}_1)
\end{array}
\end{equation}
here the term $\lbrack \cdots\rbrack^{\ast }$ denotes complex
conjugated part of the former parts. Varying $S_{\mathcal{G}\phi
B}$ with respect to $\mathcal{B}_{\mu}$ gives directly
\begin{equation} %(78)
\check{c}^{\mu \nu \lambda } (\stackrel{a}{\nabla }_{\nu}
\mathcal{G}_{\lambda }-\stackrel{a}{\nabla }_{\lambda
}\mathcal{G}_{\nu}) = -\check{c}^{\mu \nu \lambda }
\mathcal{F}_{\nu \lambda }
\end{equation}
Plugging this back into $S_{\mathcal{G}\phi \mathcal{B}}$,
eliminating $\mathcal{B}_{\mu } $ from it, gives Dirac Lagrangian
$(34)$, i.e.
\begin{equation} %(79)
\begin{array}{ll}
S_{G}= &\frac{1}{2}\lbrack (\partial _{\mu }G_{\nu })^{\ast
}i\check{c}^{\nu \mu \lambda }G_{\lambda }-G_{\nu }^{\ast
}i\check{c}^{\nu \mu \lambda }\partial _{\mu }G_{\lambda
}+m(G_{\nu }^{\ast }G^{\ast \nu }+G_{\nu }G^{\nu })\rbrack \\\noa6
& +\partial _{\mu }\lbrack \frac{1}{2}(\mathcal{G}_{\nu }^{\ast
}i\check{c}^{\nu \mu \lambda}\mathcal{B}_{\lambda
}-\mathcal{B}_{\nu }^{\ast } i\check{c}^{\nu \mu \lambda
}\mathcal{G}_{\lambda})+Z^{\mu}_1 \rbrack
\end{array}
\end{equation}
here $G^{\mu }=\mathcal{G}^{\mu }-j^{\mu }\phi $ and $A_{\mu}=0$.
The additional last term is the total divergence and does not
contribute to the equation of motion. We choose $Z_{1}^\mu$ such
that the last surface term is equal to zero. We want to notice
that elimination of  $\mathcal{B}^{\mu }$ from $S_{\mathcal{G}\phi
\mathcal{B}}$ is not so easy here; the solution of equation (78)
for $\mathcal{B}^{\mu }$ is non-trivial. However, we have not
attempted to an explicit solution of Eq.\ (78) here.

Alternatively, varying $\mathcal{G}_{\nu }$ in the parent action
$\int (d^{4}x)S_{\mathcal{G}\phi \mathcal{B}}$ yields
\begin{equation} %(80)
(m-a) \mathcal{G}^{\mu }=\frac{i}{2}\mathcal{F}_{\lambda \nu
}^{\ast }\check{c}^{\nu \lambda \mu }-ip_{+}^{\mu \nu } \partial
_{\nu }\phi ^{\ast }
\end{equation}
here the projective operator $p_{+}^{\mu \nu }=(\eta ^{\mu \nu
}+j^{\mu }j^{\nu })$. Eliminating $\mathcal{G}_{\nu }$ from
$S_{\mathcal{G}\phi \mathcal{B}}$ we obtain
\begin{equation} %(81)
\begin{array}{ll}
S_{\phi \mathcal{B}}= & \frac{1}{8(m-a)}(\check{c}^{\nu \mu
\lambda }\mathcal{F}_{\mu \lambda })(\check{c}^{\nu \rho \delta
}\mathcal{F}_{\rho \delta}) +\frac{1}{4}\mathcal{B}_{\nu }^{\ast
}i\check{c} ^{\nu \mu \lambda }\mathcal{F}_{\mu \lambda }
\\\noa6 &-\frac{1}{2(m-a)}(\partial _{\mu }\phi
)\check{c}^{\mu \nu \lambda }\mathcal{F}_{\nu \lambda }\\\noa6
&+\frac{1}{2(m-a)}(\partial _{\mu }\phi )p_{+}^{\mu \nu }(\partial
_{\nu }\phi )+\frac{i}{2}\phi ^{\ast }j^{\mu }\stackrel{-m}{\nabla
}_{\mu }\phi \\\noa6 &+\lbrack \cdots\rbrack ^{\ast
}+\partial_{\mu}(Z_2^\mu )
\end{array}
\end{equation}
here $\lbrack \cdots\rbrack ^{\ast }$ denotes complex conjugated
part of the former parts, $\mathcal{B}_{\nu }j^{\nu}=0$  and
\begin{equation} %(82)
Z_2^\mu =\lbrack Z^{\mu}_1 + \frac{i}{2} \mathcal{G}^{\ast
\mu}\phi -\frac{i}{2} \mathcal{G}^{\mu}\phi ^{\ast} \rbrack
\end{equation}
The last term in Eq.\ (81) is a surface term and does not contribute to
the equation of motion. Thus, from the parent action $\int
(d^{4}x)S_{\mathcal{G}\phi \mathcal{B}}$ we have shown that
Dirac's action $\int (d^{4}x)S_{G}$ and new action $\int
(d^{4}x)S_{\mathcal{B}\phi}$ is dual to each other; the two
actions represent the same physics (at least classically).

The corresponding equations of motion for new fields
$\mathcal{B}^{\mu }$ and $\phi$ are
\begin{equation} %(83)
\begin{array}{rcl}
\stackrel{m}{\nabla }_{\nu }\mathcal{F}^{\nu \mu }-\frac{m+a}{2}
\mathcal{F}_{\nu \lambda }^{\ast }\check{\epsilon}^{ \lambda \nu
\mu}&=&j^{\nu } \stackrel{a}{\nabla }_{\nu} \stackrel{-m}{{\nabla
}^{\mu }}\phi
\\\noa4 (\partial\partial\phi+m^2 \phi) &=& 0
\end{array}
\end{equation}
Clearly they are identical with Dirac equation (36) when equation
(80) is used.

There is a gauge freedom in the above model. The tensor
$\mathcal{F}^{\mu \nu }$ which is defined by Eq.\ (76), is
equivalent to $B^{\mu \nu}$ up to axial gauge and is invariant
against the following gauge transformation
\begin{equation} %(84)
\mathcal{B}^{\mu}\longrightarrow \mathcal{B}^{\mu}+p_{+}^{\mu \nu}
\partial_{\nu} \Theta
\end{equation}
Here $\Theta=\beta ch(-a xj)+i\beta^{\ast} sh(-a xj)$ and $\beta$
is localized only in a ``spin manifold" $M^{2+1}$ (i.e.,
$j^{\mu}\partial_{\mu}\beta=0$). The equation of motion is
invariant against the above gauge transformation. While the
corresponding Lagrangian (81) is not gauge-invariant but changes
only by a total derivative.

As mentioned in the previous section, it is convenient to use
(antisymmetric) self-dual tensor $\mathcal{H}^{[\mu \nu]}$ instead
of $\mathcal{F}^{\mu \nu }$.  On the analogy of Eqs.\ (69)(70)
\begin{equation} %(85)
\begin{array}{rl}
\mathcal{H}^{[\mu \nu]}\stackrel{\mbox{\scriptsize
def.}}{=}&\frac{i}{2(m-a)} c^{\mu \lambda \rho \nu
}\mathcal{F}_{\lambda \rho} =\frac{i}{(m-a)}(\mathcal{F}^{\mu \nu
}-\frac{i}{2} \epsilon^{\mu \nu \lambda \rho}\mathcal{F}_{\lambda
\rho}) \\\noa6 \mathcal{G}^{\ast \mu }=& -\mathcal{H}^{[\mu
\nu]}j_{\nu}+\frac{i}{m-a}p_{+}^{\mu \nu }
\partial _{\nu }\phi
\end{array}
\end{equation}

With these notations the Lagrangian (81) takes the following form
\begin{equation} %(86)
\begin{array}{ll}
S_{\phi \mathcal{B}}= & \frac{m-a}{8}\mathcal{H}^{[\mu
\nu]}\mathcal{H}_{[\mu \nu]}+i \mathcal{H}^{[\mu \nu]}j_{\nu}
(\partial_{\mu}\phi-\frac{i(m-a)}{2}\mathcal{B}^{\ast}_{\mu})
\\\noa6 & +\frac{1}{2(m-a)}(\partial _{\mu }\phi
)p_{+}^{\mu \nu }(\partial _{\nu }\phi )+\frac{i}{2}\phi ^{\ast
}j^{\mu }\stackrel{-m} {\nabla }_{\mu }\phi \\\noa6 &+\lbrack
\cdots\rbrack ^{\ast } +\partial _{\mu }(Z^{\mu}_2)
\end{array}
\end{equation}
The equations of motion read
\begin{equation} %(87)
\begin{array}{rll}
\stackrel{-m}{\nabla }_{\nu } \mathcal{H}^{[\nu \mu]}&=&
\frac{i}{m-a}j^{\nu}\stackrel{a}{\nabla }_{\nu }
\stackrel{-m}{{\nabla }^{\mu }}\phi\\\noa6
\partial \partial \phi+m^2 \phi&=&0
\end{array}
\end{equation}
$\mathcal{H}_{[\mu \nu]}$ must be an invariant object. In vacuum
$(\mathcal{G}^{\mu},\phi)=(0,0)$. Thus according to Eq.\ (85), the
condition for pure gauge vacuum must be $\mathcal{H}_{[\mu
\nu]}=0$ or equivalently $\mathcal{F}^{\mu \nu }-\frac{i}{2}
\epsilon^{\mu \nu \lambda \rho}\mathcal{F}_{\lambda \rho}=0$ (not
necessary $\mathcal{F}^{\nu \mu }=0$ or $\delta
\mathcal{B}^{\mu}=p_{+}^{\mu \nu}
\partial_{\nu}\Theta$). And
this implies the self-dual property of the pure gauge vacuum.

Now consider the particle confined to motion on a plane (like Holl
effect). In other words consider the fields $\mathcal{B}^{\mu}$
and $\phi$ on an oriented three-dimensional spin manifold
$M^{2+1}$ (i.e., $j^{\nu}\partial_{\nu}\mathcal{B}^{\mu}=0$ and
$j^{\nu}\partial_{\nu}\phi=0$). If $a=0$ then the Lagrangian (81)
and equations of motion (83) take the following form
\begin{equation} %(88)
\begin{array}{ll}
S_{\phi \mathcal{B}}= & \frac{-1}{4m}(\mathcal{F}_{\mu \lambda
}\mathcal{F}^{\mu \lambda} -m\mathcal{B}_{\nu }^{\ast
}\check{\epsilon} ^{\nu \mu \lambda }\mathcal{F}_{\mu \lambda
})\\\noa6 & +\frac{1}{2m}[(\partial _{\mu }\phi )p_{+}^{\mu \nu
}(\partial^{\nu }\phi )-m^2 \phi \phi] \\\noa6 &+\lbrack
\cdots\rbrack ^{\ast }+\partial_{\mu}[(\cdots)^\mu]
\end{array}
\end{equation}
\begin{equation} %(89)
\begin{array}{rcl}
\partial_{\nu }\mathcal{F}^{\nu \mu }+\frac{m}{2}
\check{\epsilon}^{\mu \nu \lambda}\mathcal{F}_{\nu \lambda }^{\ast
}&=&0\\\noa6
(\partial\partial\phi+m^2 \phi) &=& 0
\end{array}
\end{equation}
Notice: there is no interaction between $(\mathcal{B}_1)_{\mu}$,
$(\mathcal{B}_2)_{\mu}$ and $\phi$  (here
$\mathcal{B}_{\mu}=(\mathcal{B}_1)_{\mu}+i(\mathcal{B}_2)_{\mu}$).
In fact, the interaction term in the Lagrangian has become a
surface-term and it does not contribute to the equations of
motion. This model does not contain the symmetric Jordan
structure. Thus the pure gauge vacuum must be denoted by
$\mathcal{F}^{\mu \nu }=0$ or equivalently $\delta
\mathcal{B}^{\mu}=p_{+}^{\mu \nu}
\partial_{\nu}\Theta$. We
conclude that on the spin-manifold the 3d-vector part of Dirac
field $\mathcal{G}_{\mu}$ is dual-equivalent to 3d topologically
massive gauge field$^{[2]}$.

\section{Non-integrable exponential factor.}

The concept of ``non-integrable phase factor" was introduced and
has studied by H.\ Weyl$^{[18]}$ (1929), Dirac$^{[19,20]}$ (1931),
C.N.\ Yang$^{[21]}$, etc. Now, physicists know that the effect of an
electromagnetic potential is to introduce a non-integrable phase
in the wave function of the potential-free particle $\Psi={\Psi}_0
\cdot e^{i\int eA_\mu dx^\mu } $. Electromagnetism is thus the
gauge-invariant manifestation of this non-integrable phase
factor$^{[21]}$. Dirac emphasized that ``non integrable phases are
perfectly compatible with all the general principles of quantum
mechanics and do not in any way restrict their physical
interpretation". We propose here an alternative way to generalize
the above idea to the massive fermion.

Let us start with the fundamental ``normed condition" for
biquaternion, i.e.\ Eq.(2). For our biquaternion $G^{\mu}$ it means
that
\begin{equation} %(90)
\begin{array}{ll}
(G^{\ast }_{\mu} G^{\ast \mu})(G_{\sigma} G^{\sigma})=&(G^{\ast
}_{\sigma} c^{\sigma \mu \lambda} G_{\lambda})\eta_{\mu \nu}
(G^{\ast }_{\rho} c^{\rho \nu \tau}G_{\tau} )\\\noa4 (G^{\ast
}_{\mu} G^{\ast \mu})(G_{\sigma} G^{\sigma})=-&(G^{\ast }_{\sigma}
\check{c}^{\sigma \mu \lambda} G_{\lambda})\eta_{\mu \nu} (G^{\ast
}_{\rho} \check{c}^{\rho \nu \tau}G_{\tau})
\end{array}
\end{equation}
From Dirac's Lagrangian (34) we know that $e\pi^{\mu}=eG^{\ast
}_{\rho} c^{\rho \mu \sigma}G_{\sigma}$ is a charge current, and
from physical point of view a charged particle must be massive.
According to the principle of special relativity, for massive
fermion $\pi_{\mu}\pi^{\mu} \neq 0$ ! Thus from the above equation
we know that for massive fermion $(G_{\sigma} G^{\sigma}) \neq 0$
and thus
\begin{equation} %(91)
G^{\ast }_{\mu}= \left(\frac{G^{\ast }_{\rho}c^{\rho \nu
\tau}G_{\tau}}{G_{\sigma} G^{\sigma}}\right)c^{\mu \nu \lambda}
G_{\lambda}=-\left(\frac{G^{\ast }_{\rho}\check{c}^{\rho \nu
\tau}G_{\tau}}{G_{\sigma} G^{\sigma}}\right)\check{c}^{\mu \nu \lambda}
G_{\lambda}
\end{equation}
Let $G^\mu=\mathcal{R}^{\mu}+\mathcal{L}^{\mu}$ (see eq.(39)),
then the above equation becomes
\begin{equation} %(92)
\begin{array}{rll}
\mathcal{R}^{\ast \mu}&=\check{c}^{\mu \nu \lambda }K_\nu
\mathcal{L}_{\lambda}&=\check{c}^{\mu \nu \lambda }(K^{+}_\nu)
\mathcal{L}_{\lambda}
\\\noa4
-\mathcal{L}^{\ast\mu}&=\check{c}^{\mu \nu \lambda }K^{\ast}_\nu
\mathcal{R}_{\lambda}&=\check{c}^{\mu \nu \lambda
}(K^{-}_\nu)^{\ast} \mathcal{R}_{\lambda}
\end{array}
\end{equation}
here $\mathcal{R}^{\mu}\mathcal{L}_{\mu}\neq 0 $ and
\begin{equation} %(93)
\begin{array}{c}
\displaystyle K_{\nu}=\frac{\mathcal{R}^{\ast}_{\mu} c^{\mu \nu \lambda}
\mathcal{R}_{\lambda}}{2\mathcal{L}_{\rho}\mathcal{R}^{\rho}}
+\frac{\mathcal{L}^{\ast}_{\mu} c^{\mu \nu \lambda}
\mathcal{L}_{\lambda}}{2(\mathcal{L}_{\rho}\mathcal{R}^{\rho})^{\ast}}=K^{+}_\nu
+K^{-}_\nu
\\\noa6
K_{\nu} K^{\nu}=1
\end{array}
\end{equation}
Thus the Dirac equation (40) can be rewritten in the following
form
\begin{equation} %(94)
\begin{array}{ll}
\check{c}^{\mu \nu \lambda }[\partial _{\nu }+i(eA_{\nu
}+mK_{\nu})]^{\ast }\mathcal{R}_{\lambda}&=0
 \\\noa4
\check{c}^{\mu \nu \lambda }[\partial _{\nu }+i(eA_{\nu
}+mK_{\nu})]\mathcal{L}_{\lambda}&=0
\end{array}
\end{equation}
It is important to notice again that a charged particle must be
massive. In the above formalism the mass $m$ was introduced in the
similar way as the charge $e$. It is mathematically beautiful and
physically natural. This formalism leads to the concept of
non-integrable exponential factor. The equation of motion (94) can
be rewritten in the following form
\begin{equation} %(95)
\begin{array}{l}
\check{c}^{\mu \nu \lambda }\partial _{\nu }
\mathcal{R}_{0\lambda}=0 \\\noa3
\check{c}^{\mu \nu \lambda }\partial _{\nu
}\mathcal{L}_{0\lambda}=0
\end{array}
\end{equation}
 where
\begin{equation} %(96)
\begin{array}{l}
\mathcal{R}_{0}^{\mu}=\mathcal{R}^{\mu}\cdot e^{-i\int_{x_0}^{x}
(eA_{\nu}+mK_{\nu})^{\ast}dx^{\nu}}\\\noa3
\mathcal{L}_{0}^{\mu}=\mathcal{L}^{\mu}\cdot e^{i\int_{x_0}^{x}
(eA_{\nu}+mK_{\nu})dx^{\nu}}
\end{array}
\end{equation}
The mapping $(\mathcal{R}^{\mu},\mathcal{L}^{\mu})\rightarrow
(\mathcal{R}_{0}^{\mu},\mathcal{L}_{0}^{\mu}) $ includes a
non-integrable exponential factor, it depends on the path of
integration from some initial point $x^0$ to $x^{\mu}$. Our
formalism can be considered as generalization of the ``theory of
non-integrable phases" which was introduced and studied by H.\
Weyl, P.A.M.\ Dirac, and C.N.\ Yang.

It is important to notice that although in Eqs.\ (93) and (96)
$K_\mu$ is dependent on $(\mathcal{R}^{\mu},\mathcal{L}^{\mu})$,
but there is an inverse correspondence between
$(\mathcal{R}^{\mu},\mathcal{L}^{\mu})$ and
$(\mathcal{R}_{0}^{\mu},\mathcal{L}_{0}^{\mu}) $. Using Eqs.(93)
and (96) we obtain
\begin{equation} %(97)
\frac{\mathcal{R}^{\ast}_{0\mu} c^{\mu \nu \lambda}
\mathcal{R}_{0\lambda}}{2\mathcal{L}_{0\rho}
\mathcal{R}_{0}^{\rho}} +\frac{\mathcal{L}^{\ast}_{0\mu} c^{\mu
\nu \lambda} \mathcal{L}_{0\lambda}}{2(\mathcal{L}_{0\rho}
\mathcal{R}_{0}^{\rho})^{\ast}} =\frac{\mathcal{R}^{\ast}_{\mu}
c^{\mu \nu \lambda} \mathcal{R}_{\lambda}}{2\mathcal{L}_{\rho}
\mathcal{R}^{\rho}} +\frac{\mathcal{L}^{\ast}_{\mu} c^{\mu \nu
\lambda} \mathcal{L}_{\lambda}}{2(\mathcal{L}_{\rho}
\mathcal{R}^{\rho})^{\ast}}=K^\nu
\end{equation}
We find that after mapping $(\mathcal{R}^{\mu},\mathcal{L}^{\mu})
\mapsto(\mathcal{R}^{\mu}_0,\mathcal{L}^{\mu}_0) $ the unit
(complex) vector $K_\mu$ is \textit{form-invariant}. In other
words, unit (complex) vector
$K_\mu(\mathcal{R}^{\mu}_0,\mathcal{L}^{\mu}_0)$ is the same
function of its arguments
$(\mathcal{R}^{\mu}_0,\mathcal{L}^{\mu}_0)$ as the original
$K_\mu(\mathcal{R}^{\mu},\mathcal{L}^{\mu})$ was of \textit{its}
arguments $(\mathcal{R}^{\mu},\mathcal{L}^{\mu})$.

Thus equation (96) now reads
\begin{equation} % (98)
\begin{array}{l}
\mathcal{R}_{0}^{\mu}\cdot
e^{i\int(eA_{\nu}+mK_{\nu})^{\ast}dx^{\nu}}=\mathcal{R}^{\mu}\\
\mathcal{L}_{0}^{\mu}\cdot
e^{-i\int(eA_{\nu}+mK_{\nu})dx^{\nu}}=\mathcal{L}^{\mu}
\end{array}
\end{equation}
here $K_{\mu}$ is the function of $\mathcal{R}_{0}^{\mu}$ and
$\mathcal{L}_{0}^{\mu}$. This means that the mapping
$(\mathcal{R}^{\mu},\mathcal{L}^{\mu}) \rightarrow
(\mathcal{R}^{\mu}_0,\mathcal{L}^{\mu}_0) $ is invertible.
$(\mathcal{R}^{\mu}_0,\mathcal{L}^{\mu}_0)$ (which satisfies the
massless uncharged Dirac equation) is completely defines
$(\mathcal{R}^{\mu},\mathcal{L}^{\mu})$ (which satisfies the
massive Dirac equation) and {\it vice versa}. Under dual
transformation $(\mathcal{R}^{\mu},\mathcal{L}^{\mu})
\leftrightarrow (\mathcal{R}^{\mu}_0,\mathcal{L}^{\mu}_0) $ the
unit (complex) vector $K_\mu$ is unchanged! Although at the level
of equation of motion two models are dual-equivalent. But for
physicist it is easier, at the beginning, to work with more
symmetric massless unchanged Dirac equation.

Up to day physicists consider Dirac field as a spinor field. Thus,
it is worth while to rewrite the above result in a more familiar
spinor form.

\textbf{Theorem:} Physical spinor field $\Psi $ which satisfies
massive charged Dirac equation (i.e.\ all solutions of massive
Dirac equation) can be expressed in the following form
\begin{equation} % (99)
\Psi \stackrel{d}{=}\Psi _0\cdot e^{i\int (eA_\mu -mK_\mu )dx^\mu}
\end{equation}
here $\Psi _0=R_0+L_0$ satisfies massless (uncharged) Dirac
equation and $\bar{\Psi}_0\Psi_0 \neq 0$. The unit vector
\begin{equation}  %(100)
K_\mu =\frac{\bar{R}_0\gamma _\mu R_0}{2%
\bar{R}_0L_0}+\frac{\bar{L}_0\gamma _\mu
L_0}{2\bar{L}_0R_0}
=\frac{\bar{R}\gamma _\mu R}{2%
\bar{R}L}+\frac{\bar{L}\gamma _\mu L}{2\bar{L}R}
\end{equation}
\begin{equation}  %(101)
\begin{array}{c}
K_\mu \gamma^\mu R=L \\
K_\mu \gamma^\mu L=R
\end{array}
\end{equation}
(Remark: equations (24) can be considered as a special case of the above
equations.)

It is important to notice that in general $K_\mu
=[\mbox{Re}\,(K_\mu )+i\mbox{Im}\,(K_\mu )]$ is complex.
\begin{equation} %(102)
i\mbox{Im}\,(K_\mu ) = \frac{(\bar{\Psi}_0\gamma _\mu \gamma
^5\Psi_0) (\bar{\Psi}_0\gamma ^5\Psi_0 )}{(\bar{\Psi}_0
\gamma _\nu \Psi_0)(\bar{\Psi}_0 \gamma ^\nu \Psi_0)}
\end{equation}
\begin{equation} % (103)
\mbox{Re}\,(K_\mu ) =\frac{(\bar{\Psi}_0 \gamma _\mu \Psi_0)
(\bar{\Psi}_0 \Psi_0 )}{(\bar{\Psi}_0 \gamma _\nu
\Psi_0)(\bar{\Psi}_0 \gamma ^\nu \Psi_0)}
\end{equation}
The imaginary part Im$\,(K_\mu )$ is associated with the axial
current $\pi^{\mu}_5$ and corresponds to the ``scale factor'' of a
spinor. While the real part Re$\,(K_\mu )$ is associated with vector
current $\pi^{\mu}$ and corresponds to the phase of the spinor.

Up to date, the geometrical and physical interpretation for unit
vector $K_\mu$ is not clear for us. However for physicists, the
plane-wave solution is the most important solution in the quantum
field theory. In this special case, $m\,\mbox{Re}\,(K_\mu)$ is
nothing but the energy-momentum of massive Dirac particle.

A particularly tantalizing result by Dirac, in study of the
concept of ``non-integrable exponential-phase", concerns his
monopoles. As is well known, he showed that within quantum mechanics
monopole strength has to be quantized, but the quantization does
not arise from a quantal eigenvalue problem. Rather quantization
is enforced by requirement that the exponential-phase in the wave
function must be quantized. The interest of the theory of magnetic
poles is that it forms a natural generalization of the usual
electrodynamics and it leads to the quantization of electricity.
Dirac emphasized that ``the quantization of electricity is one of
the most fundamental and striking features of atomic physics, and
there seems to be no explanation for it apart from the theory of
poles"$^{[20]}$.

We propose here an alternative way to generalize Dirac's idea to
the massive fermion. Consider the phenomena confined to motion on
a \textbf{plane}, like Holl effect. In the special coordinate
system (27) and (28), let
\begin{equation} %(104)
G^\mu_0=\frac{a}{2} \left( \begin{array}{c} 2i \\
b(x-iy)/r^2 \\
b(y+ix)/r^2 \\ 0
\end{array}\right)
{ } \longrightarrow { }
\Psi^{\alpha }_0 =\frac{a}{2}\left( \begin{array}{c} 2i \\
b(x+iy)/r^2 \\ 0 \\ -b(x+iy)/r^2
\end{array}
\right)
\end{equation}
here $a, b$ are real constants, $x^\mu=(t,x,y,z)$ and $r^2 =x^2
+y^2$. It is easy to prove that the above $G^\mu_0$ (and $\Psi^{\alpha
}_0(x,y)$) satisfies massless uncharged Dirac equation, and the
unit vector
\begin{equation} %(105)
K^\mu =\left(
\begin{array}{c}
1+b^2/(2r^2) \\
-by/r^2 \\
bx/r^2 \\
b^2/(2r^2)
\end{array}
\right)
\end{equation}
is real. $\Psi =\Psi _0\cdot e^{i\int (eA_\mu -mK_\mu )dx^\mu}$
satisfies massive Dirac equation and $\bar{\Psi} \Psi=a^2$.

Because of the single-valued nature of quantum mechanical wave
function, we can naturally conjecture that the mapping $\Psi_0
\rightarrow \Psi$ (on the plane) is single-valued, and it leads to
the requirement that: the \textit{phase} change of a wave function
$\oint_C (eA_\mu -mRe(K_\mu) )dx^\mu$ round any closed curve $C$
on the $(x,y)$ plane,  must be close to $2n\pi$ where $n$ is some
integer. This integer will be a characteristic of possible
singularity in $A_\mu$ and Re$\,(K_\mu)$. In our case it means
that
\begin{equation} %(106)
\oint_C (eA_\mu -m\,\mbox{Re}\,(K_\mu)) dx^\mu=(\oint_C eA_\mu dx^\mu)
+2{\pi}bm=2{\pi}n
\end{equation}
and it leads to the law of {\it quantization} of physical
constants. The charged particle must be massive, thus the
quantization of electricity must have close relation with the
mass. Ap to date, the existence of magnetics monopole is an open
question yet, thus the above equation will lead to the law of
quantization of constant $(bm)$ which includes the mass parameter.

\end{document}